\documentclass[12pt]{article}

\usepackage{latexsym}
\usepackage{amsthm}
\usepackage{amsmath}
\usepackage[dvips]{graphicx}
\usepackage{amssymb}

\newcommand{\be}{\begin{eqnarray}}
\newcommand{\ee}{\end{eqnarray}}

\begin{document}

\baselineskip=18pt

\setcounter{footnote}{0}
\setcounter{figure}{0}
\setcounter{table}{0}

\begin{titlepage}

\begin{center}
\vspace{.1cm}

{\Large \bf The S-Matrix in Twistor Space}

\vspace{0.1cm}

{\bf N. Arkani-Hamed$^a$, F. Cachazo$^b$, C. Cheung $^{a,c}$ and J. Kaplan$^{a,c}$}

\vspace{.1cm}

{\it $^{a}$ School of Natural Sciences, Institute for Advanced Study, Princeton, NJ 08540, USA}

{\it $^{b}$ Perimeter Institute for Theoretical Physics, Waterloo, Ontario N2J W29, CA}

{\it $^{c}$ Jefferson Laboratory of Physics, Harvard University, Cambridge, MA 02138, USA}

\end{center}

\begin{abstract}

The marvelous simplicity and remarkable hidden symmetries recently
uncovered in (Super) Yang-Mills and (Super)Gravity scattering
amplitudes strongly suggests the existence of a ``weak-weak" dual
formulation of these theories where these structures are made more
manifest at the expense of manifest locality. In this note we
suggest that in four dimensions, this dual description lives in
(2,2) signature and is naturally formulated in twistor space. We
begin at tree-level, by recasting the momentum-space BCFW recursion
relation in a completely on-shell form that begs to be transformed
into twistor space. Our transformation is strongly inspired by
Witten's twistor string theory, but differs in treating twistor and
dual twistor variables on a more equal footing; a related
transcription of the BCFW formula using only twistor space variables
has been carried out independently by Mason and Skinner. Using both
twistor and dual twistor variables, the three and four-point
amplitudes are strikingly simple--for Yang-Mills theories they are
``1" or ``-1". The BCFW computation of higher-order amplitudes can
be represented by a simple set of diagrammatic rules, concretely
realizing Penrose's program of relating ``twistor diagrams" to
scattering amplitudes. More specifically, we give a precise
definition of the twistor diagram formalism developed over the past
few years by Andrew Hodges. The ``Hodges diagram" representation of
the BCFW rules allows us to compute amplitudes and study their
remarkable properties in twistor space. For instance the diagrams
for Yang-Mills theory are topologically disks and not trees, and
reveal striking connections between amplitudes that are not manifest
in momentum space. Twistor space also suggests a new representation
of the amplitudes directly in momentum space, that is naturally
determined by the Hodges diagrams. The BCFW rules and Hodges
diagrams also enable a systematic twistorial formulation of gravity.
All tree amplitudes can be combined into an ``S-Matrix" scattering
functional which is the natural holographic observable in
asymptotically flat space; the BCFW formula turns into a simple
quadratic equation for this ``S-Matrix" in twistor space, providing
a holographic description of ${\cal N} = 4$ SYM and ${\cal N} = 8$
Supergravity at tree level. We move on to initiate the exploration
of loop amplitudes in $(2,2)$ signature and twistor space, beginning
with a discussion of their IR behavior. We find that the natural
pole prescriptions needed for transformation to twistor space make
the amplitudes perfectly well-defined objects, free of IR
divergences. Indeed in momentum space, the loop amplitudes so
regulated vanish for generic momenta, and transformed to twistor
space,  are even simpler than their tree-level counterparts: the
full 4-pt one-loop amplitudes in ${\cal N} = 4$ SYM are simply equal
to ``1" or ``0"! This further supports the idea that there exists a
sharply defined object corresponding to the S-Matrix in (2,2)
signature, computed by a dual theory naturally living in twistor
space.

\end{abstract}

\bigskip
\bigskip

\end{titlepage}

\section{Towards A Dual Theory of the S-Matrix}

The past two decades have seen a growing realization that scattering amplitudes in gauge theory and gravity exhibit amazing properties that are invisible in the usual local formulation of field theory \cite{reviews}, ranging from the stunning simplicity of MHV amplitudes \cite{ParkeTaylor} to the recent discovery of dual-superconformal invariance \cite{dual,dualinloops} to the surprisingly good UV behavior of (super)gravity amplitudes \cite{gravgood, gravgood2, reviewsugra}. These discoveries strongly suggest the existence of a holographically dual formulation of (Super) Yang-Mills and (Super) Gravity that should make these remarkable properties manifest at the expense of manifest locality. In this note and in a subsequent paper \cite{HolS}, we suggest that in four dimensions this sought-after dual description should live in (2,2) signature and finds a natural description in twistor space.

After a quick introduction to the kinematical aspects of (2,2) twistor space relevant to our discussion, we show that the BCFW recursion relations for tree-level amplitudes \cite{BCFW,BCF,RRgeneral,generalproof}, when cast in their most natural on-shell form, ask to be fourier-transformed into twistor space, now revealed as the natural home of the BCFW formalism. The three and four point functions are amazingly simple in twistor space, and the the BCFW computation of higher-order amplitudes can be represented by a simple set of diagrammatic rules. This concretely realizes Penrose's program, dating from the 1970's, of relating
what he called ``twistor diagrams" to scattering amplitudes \cite{Penrose, HodgesOld,moderntwistors}. In recent years the twistor diagram formalism has been vigorously developed by Andrew Hodges~\cite{Hodges}, and we make very direct contact with his work. Indeed our diagrammatic rules give a precise definition of Hodges' diagrams. His diagrams are associated with contour integrals in complex twistor space, but the choice of the contour of integration is non-trivial and has not yet been made systematic; our construction in (2,2) signature involves real integrals and can be thought of as specifying at least one correct contour of integration. The ``Hodges diagram" representation of the BCFW rules is quite powerful, and allows us to compute the amplitudes and study their properties in twistor space. For instance the diagrams for Yang-Mills theory are topologically disks rather than trees, which is strongly suggestive of an underlying open string theory. The Hodges diagrams also reveal connections between the scattering amplitudes that are not manifest in momentum space. The structure of twistor space amplitudes also suggest a novel way of writing amplitudes directly in momentum space--which we call the ``link representation"--and we show in some examples how this can be read off directly from the Hodges diagrams. It should also be emphasized that the BCFW rules and Hodges diagrams can be used to initiate a systematic study of gravity in twistor space!

Our transformation to twistor space is clearly very strongly inspired by Witten's 2003 twistor string theory~\cite{WittenTwistor}, but differs in treating twistor and dual twistor variables on an equal footing. While our work was in progress, we learned of independent work by Lionel Mason and David Skinner \cite{LionelDavid}, who write the BCFW recursion relations using only twistor variables. Our formulations are related to each other by full twistor-space fourier transforms, as described in more detail in \cite{LionelDavid}. Our formalism (for the case of Yang-Mills) also appears to be closely related to Witten's 1978 twistorial formulation of the classical equations of motion for Yang-Mills theory~\cite{Witten78}. A feature of both constructions is that, even at tree-level, maximally supersymmetric theories are much more natural in twistor space than their non-supersymmetric counterparts.

All tree amplitudes can be combined into an ``S-Matrix" scattering functional which is the natural holographic observable in asymptotically flat space;  the BCFW formula turns into a strikingly simple quadratic equation for this ``S-Matrix" in twistor space. This equation can be thought of as providing a completely holographic definition of ${\cal N} = 4$ Super-Yang-Mills and ${\cal N} = 8$ Supergravity at tree level.

We next begin a study of the properties of loop amplitudes in
$(2,2)$ signature and twistor space. Of course physics in $(2,2)$
signature is neither causal nor unitary in any standard sense;
there is no good physical interpretation of tree amplitudes, not
to speak of an understanding of what loops are and why they are
needed. Nonetheless, we press ahead with an exploratory attitude,
and examine the properties of loop integrals in split signature.
In the usual Lorentzian signature, a common complaint about even
talking about scattering amplitudes for massless particles beyond
tree-level is that ``they don't exist", due to IR divergences
which have to be regulated by using e.g. dimensional
regularization. We therefore  begin with an exploration of the IR
structure of loop amplitudes in split signature, which turns out
to be more interesting than its Lorentzian counterpart. We find
that, regulating all relevant poles with a principle value
prescription natural both for (2,2) signature and transformation
to twistor space, the loop amplitudes are perfectly well-defined
objects, free of IR divergences. Indeed in momentum space, the
loop amplitudes vanish for generic momenta, and transformed to
twistor space, are even simpler than their
tree-level counterparts; for instance the full 4-pt one-loop
amplitudes in ${\cal N} = 4$ SYM are simply equal to ``1" or ``0"!
This adds further fuel to the idea that there is a perfectly
well-defined object corresponding to the S-Matrix living in (2,2)
signature, computed by a dual theory naturally formulated in
twistor space.

Our purpose in this rather telegraphic note is to motivate the transformation to twistor space and introduce some of the relevant formalism. In our next paper \cite{HolS}, we will describe this formalism and its physical content in much more detail, and go on to discuss further developments taking us beyond Hodges diagrams and the BCFW formalism, closer towards a real dual theory of the S-Matrix.

\section{Twistor Space Kinematics}

There are several motivations for studying scattering amplitudes in twistor space.
An early motivation was that twistor space allows us to talk about the free asymptotic states, associated with linearized classical solutions, in the simplest way. It also allows us to realize the symmetries of scattering amplitudes in the most transparent setting. As we will see in the next section, when appropriately formulated in a completely on-shell fashion in (2,2) signature, the BCFW recursion relations also demand a transcription into twistor space. But before getting there we begin with a quick review of some twistor-space basics~\cite{Penrose}, to set notation and highlight a few essential points.

Consider the scattering amplitude
\begin{equation}
M(\lambda_i,\tilde{\lambda}_i;h_i) =  {\cal M}(\lambda_i,\tilde{\lambda}_i;h_i) \delta^4(\sum_i \lambda_i \tilde \lambda_i)
\end{equation}
for a set of particles labeled by $i$, with helicity $h_i$, and including the momentum-conserving delta function. Under the little group action $\lambda_i \to t_i \lambda_i, \, \, \tilde{\lambda}_i \to t_i^{-1} \tilde{\lambda}_i$
we have
\begin{equation}
M(\lambda_i,\tilde{\lambda}_i;h_i) \to t_i^{-2 h_i} M(\lambda_i,\tilde{\lambda}_i;h_i)
\end{equation}
Now let us suppose we are in $(2,2)$ signature so that the $\lambda,\tilde{\lambda}$ are independent real spinors. To go to twistor space for a given particle we simply fourier transform with respect to the $\lambda$ variable, while going to the dual twistor space is accomplished by fourier transforming with respect to $\tilde{\lambda}$. Thus, we have
\begin{equation}
M(\cdots,W,\cdots) = \int d^2 \lambda e^{i \tilde \mu^a \lambda_a} M(\cdots, \lambda, \cdots), \, M(\cdots,Z,\cdots) = \int d^2 \tilde{\lambda}  e^{i \mu^{\dot{a}} \tilde{\lambda}_{\dot{a}}} M(\cdots, \tilde \lambda, \cdots)
\end{equation}
where
\begin{equation}
W_A = \left(\begin{array}{c} \tilde{\mu} \\ \tilde{\lambda} \end{array} \right), \, Z^A = \left(\begin{array}{c} \lambda \\ \mu \end{array} \right)
\end{equation}
Note that we are using a convention where $\mu$ and $\tilde \mu$ variables have dotted and un-dotted lorentz indices, in the opposite way as $\lambda, \tilde \lambda$.
We use lower and upper $A$ indices on $Z_A,W^A$ to emphasize the fact that the full conformal group acts very simply as $SL(4,R)$ transformations on these four-vectors. Conformal invariants can be built just out of $W$'s using the $\epsilon$ tensor as $\epsilon^{ABCD} W_A W_B W_C W_D$. With both $W$'s and $Z$'s we have the natural invariant
\begin{equation}
W_A Z^A \equiv W \cdot Z = \tilde \mu \lambda - \mu \tilde \lambda
\end{equation}
while objects that are only invariant under the poincare subgroup of the full conformal group are made using the ``infinity twistors" $I_{AB}, I^{AB}$
\begin{equation}
W_1 I W_2 \equiv [\tilde \lambda_1 \tilde \lambda_2], \, Z_1 I Z_2 \equiv \langle \lambda_1 \lambda_2 \rangle
\end{equation}
Furthermore, under the little group action, the $Z,W$ variables
transform homogeneously as $W \to t^{-1} W, Z \to t Z$ so that the
amplitudes are naturally projective objects; for particles of spin
$s$ we have
\begin{eqnarray}
M(tW;+) = t^{2(s - 1)} M(W;+) &, &  M(tZ;-) = t^{2(s - 1)} M(Z;-) \\
M(tW;-) = t^{-2(s+1)} M(W;-) &,& M(tZ;+) = t^{-2(s+1)} M(Z;+)
\end{eqnarray}
Thus the amplitudes should be thought of as ``densities" on $\mathbb{RP}^3$ with appropriate projective weights, though we will refer to them more colloquially as ``functions" on $\mathbb{RP}^3$. We can go back and forth between the $Z$ and $W$ representations by a full $d^4Z$ or $d^4W$ fourier transformation (or ``twistor transform")
\begin{equation}
M(\cdots, W, \cdots) = \int d^4 Z e^{i Z \cdot W} M(\cdots, Z, \cdots)
\end{equation}
So far our discussion has been non-supersymmetric, the maximally supersymmetric extension is completely straightforward; indeed as we will see, in many ways amplitudes virtually beg to live in maximally supersymmetric twistor space. The amplitudes for ${\cal N} = 4$ SYM and ${\cal N} = 8$ SUGRA are most naturally described by labeling the external states by Grassmann coherent states $|\eta_I \rangle$ or $|\tilde \eta^I \rangle$, which are eigenstates of $Q_{\alpha I}$ and $\tilde{Q}_{\dot{\alpha}}^I$~\cite{Nair, onshellsuperspace, simplest}; following the conventions of \cite{simplest} $|\eta = 0 \rangle = |- s \rangle$ is the negative helicity gluon or graviton, $|\tilde \eta = 0 \rangle = |+ s\rangle$ is the positive helicity gluon or graviton.
Thus we can label a given external state by either $\eta$ or $\tilde \eta$, and amplitudes are completely smooth objects $M(\cdots, \lambda,\tilde \lambda, \eta, \cdots)$ or $M(\cdots, \lambda, \tilde \lambda, \tilde \eta, \cdots)$. As a convention, when the external state is labeled by $\tilde \eta$ we will fourier-transform with respect to $\lambda$, and when it is labeled by $\eta$ we will fourier transform with respect to $\tilde \lambda$, giving us super-twistor space variables
\begin{equation}
{\cal W} = \left(\begin{array}{c} W_A \\ \tilde \eta^I \end{array} \right), \, {\cal Z} = \left(\begin{array}{c} Z^A \\ \eta_I \end{array} \right)
\end{equation}
Superconformal transformations are then just the natural supersymmetric extension of the $SL(4,R)$ transformations in the non-supersymmetric case.

Under the little group these supersymmetric amplitudes then have the same weights whether in the $W$ or $Z$ representation:
\begin{equation}
M(t {\cal W}) = t^{2 (s - 1)} M({\cal W}), \, M(t {\cal Z}) = t^{2(s - 1)} M({\cal Z})
\end{equation}
The difference with the non-supersymmetric formula comes from the extra scaling of the Grassmann $\eta, \tilde \eta$ variables under $\tilde \eta \to t \tilde \eta$, $\eta \to t \eta$.
Thus amplitudes in ${\cal N} = 4$ SYM are functions of weight zero on $\mathbb{RP}^{3|4}$, and amplitudes in ${\cal N} = 8$ SUGRA are functions of weight 2 on $\mathbb{RP}^{3|8}$. The super-twistor transform takes us from the ${\cal W}$ to the ${\cal Z}$ representation:
\begin{equation}
M({\cal W}) = \int d^{4|{\cal N}} {\cal Z} \,  e^{i {\cal Z} \cdot {\cal W}} M({\cal Z})
\end{equation}
The analogs of the invariants in the non-SUSY case are
\begin{equation}
{\cal W} \cdot {\cal Z} = W \cdot Z + \eta \cdot \tilde \eta
\end{equation}
while the invariants made with infinity twistors are unaltered
\begin{equation}
{\cal W}_1 I {\cal W}_2 = W_1 I W_2, \, {\cal Z}_1 I {\cal Z}_2 = Z_1 I Z_2
\end{equation}

Finally, with manifest ${\cal N} = 4$ or ${\cal N} = 8$ supersymmetry, the parity invariance of Super-Yang-Mills and Supergravity is obscured, since we have to pick an $\eta$ or $\tilde \eta$ variable to label the particles. Parity invariance is the  non-trivial statement that if we e.g. fourier transform all the $\eta$'s to $\tilde \eta$'s and vice-versa, and also exchange $\lambda$ with $\tilde \lambda$, the amplitude is unchanged! Transcribed into twistor space, it is the statement that
\begin{equation}
\int d^{4|{\cal N}} {\cal W}_i d^{4|{\cal N}} {\cal Z}_J e^{i {\cal W}_i \cdot {\cal Z}_i} e^{i {\cal Z}_J \cdot {\cal W}_J} M({\cal W}_i,{\cal Z}_J) = M({\cal W}_i \to {\cal Z}_i,{\cal Z}_J \to {\cal W}_J)
\end{equation}
(and in Yang-Mills theory, we also have to send $T^a \to -T^{a T})$. We emphasize that the function $M$ appearing on the right hand sign is {\it the same function} as appears on the left hand side, and not merely the amplitude in the $({\cal Z}_i,{\cal W}_J)$ basis.

\section{Transforming to Twistor Space and Back}

Let us begin our acquaince with amplitudes in twistor space by
looking at some simple examples where the explicit fourier
transformation from momentum space can be easily carried out. In
2003 Witten made the fascinating observation that gauge theory
amplitudes have remarkable properties in twistor
space~\cite{WittenTwistor}, fourier-transforming with respect to
e.g. all the $\tilde \lambda$ variables and so using only $Z$
twistor variables. This is useful to highlight the simplicity of the
MHV amplitudes, and quite naturally led to the MHV-based CSW
recursion relations~\cite{CSW}. By contrast we will not commit to
any particular choice of $W$'s and $Z$'s for the external particles;
indeed we will let the amplitudes themselves guide us to the basis
where they look simplest. However as we will see, the BCFW recursion
relations directly motivate a transformation into twistor space
where one of the BCFW particles is transformed to the $W$
representation and the other to the $Z$ representation, and we will
very generically be looking at amplitudes with a mixture of $W$'s
and $Z$'s. We are immediately rewarded for doing this by looking at
the three and four-particle amplitudes, which look incredibly simple
in a mixed $W/Z$ representation. We will then proceed to transform
back from twistor space to momentum space; the obvious way to
transform back does not trivially invert the first fourier
transformation, and we thereby  obtain a new representation of these
very familiar amplitudes back in momentum space! For reasons that
will soon become clear we call this the ``link" representation;
quite remarkably {\it all} tree amplitudes can be expressed in this
form \cite{HolS}.

\subsection{Three Particle Amplitudes}

The three-particle amplitude is a fundamental object, whose form is completely dictated by the poincare symmetries. We will shortly perform the explicit fourier-transformations to determine its form in twistor space, but since the result should be completely determined by symmetries it is also instructive to determine it directly in twistor language. Let us start with the $M^{++-}$ amplitude in Yang-Mills. If we use the $W_1,W_2, Z_3$ representation, $M^{++-}$ should simply have weight zero under independent rescalings of $W_{1,2}$ and $Z_3$. The simplest function with this property would clearly be $M^{++-} = 1$! Going back to momentum space, this corresponds to an object with very singular support at zero momentum
\begin{equation}
1 \to \delta^2(\lambda_1) \delta^2(\lambda_2) \delta^2(\tilde \lambda_3)
\end{equation}
which however of course still does conserve momentum and has the
correct little group properties! Thus, the commonly made statement
that the three-particle amplitude is fully determined by poincare
invariance actually assumes that such singular contributions are
absent; it is amusing that twistor space allows us to expose these
peculiar objects in a simple way. We will indeed find that such
strange objects arise very naturally in the computation of loop
amplitudes in $(2,2)$ signature, but they clearly don't correspond
to what we're interested in at tree-level. What we need are
non-trivial functions of the available invariants, which are $W_1
\cdot Z_3, W_2 \cdot Z_3$ and $W_1 I W_2$.  The object
corresponding to the usual Yang Mills 3pt function turns out to be
the next simplest choice:
\begin{equation}
\label{threepoint}
M_{YM}^{++-}(W_1,W_2,Z_3) = {\rm sgn} (W_1 I W_2)  \, {\rm sgn} (W_1 \cdot Z_3) \, {\rm sgn} (W_2 \cdot Z_3)
\end{equation}
where sgn$\;x$ is the sign of $x$. The ``signs" are to be expected
given that the amplitudes are naturally projective objects; note
that every $W$ are $Z$ must appear an even number of times in order
for the amplitude to have zero weight under rescaling by negative
numbers. It is very easy to see that the object above uniquely
satisfies all the necessary conditions. This is an amazingly simple
object--the three-point function in twistor space take the values
``1" and ``-1"!

Let us in particular highlight the presence of the sgn$(W_1 I W_2)$ term; it must be there for the amplitude to have the correct projective weight; it also ensures that the amplitude has the correct statistics under exchanging $1 \leftrightarrow 2$ (with the extra minus sign arising from this being a color stripped amplitude). However its presence is surprising, since we might have expected the scattering amplitude in Yang-Mills theory to be conformally invariant, and thus in twistor space to only depend on the $SL(4,R)$ invariants $W_{1,2} \cdot Z_3$, and not on the terms with Infinity twistors that only preserve the poincare symmetry. What we have just seen is that this expectation is false: the scattering amplitudes are not exactly manifestly conformally invariant! As we will see in the explicit fourier transformation in a moment, technically this arises because in the transform to twistor space, the fourier integrals needs $i \epsilon$-type regularization, and these are not conformally invariant. Note also that the non-invariance is of a mild sort; the ``sgn" term only changes it's value and reveals its breaking of conformal invariance at singular momentum configurations where $[\tilde \lambda_1 \tilde \lambda_2 ] = 0$. Thus, for small variations around generic momenta, the amplitude is conformally invariant, but large conformal transformations (and in particular inversions) detect the breaking of conformal invariance. It would be very nice to find a more physical explanation for this breaking of conformal invariance; perhaps it has to do with the fact that the scattering process does, after all, distinguish ``infinity" from the origin, since the asymptotic states are at infinity. At any rate, these ``infinity twistor sign" terms are very important and will appear everywhere in our analysis.

It is straightforward to do the direct fourier transformation of the
three particle amplitude from momentum space to twistor space. The
momentum space amplitude is
\begin{equation}
M^{++-} = \frac{[12]^3}{[13][23]}
\delta^4\left(\lambda_1 \tilde \lambda_1 + \lambda_2 \tilde \lambda_2
+ \lambda_3 \tilde \lambda_3\right) =
\frac{[12]^3}{[13][23]} \int d^4 X_{a \dot{a}} e^{i X
(\sum_i \lambda_i \tilde \lambda_i)}
\end{equation} and we will fourier transform with respect to $\lambda_1, \lambda_2$ and $\tilde \lambda_3$. The $\lambda_1,\lambda_2$ transforms are trivial since the only dependence on these is through the momentum
$\delta$ function, and we are left with
\begin{equation}
\label{threepoint} M^{++-}(W_1,W_2,Z_3) = [12]^{3} \int d^4 X
\delta^2(\tilde \mu_1 + X \tilde \lambda_1) \delta^2(\tilde \mu_2 + X \tilde
\lambda_2) \int d^2 \tilde \lambda_3 \frac{e^{i \tilde \lambda_3
(\mu_3 + X \lambda_3)}}{[13][23]}
\end{equation}
We will now perform the $\tilde \lambda_3$ integral by expanding
\begin{equation}
\tilde \lambda_3 = a_1 \tilde \lambda_1 + a_2 \tilde \lambda_2, \,
\end{equation}
Note that
\begin{equation}
d^2 \tilde \lambda_3 = |[12]| da_1 da_2 = [12] {\rm sgn}[12] da_1 da_2
\end{equation}
where we highlight the sgn$[12]$ term that is there because of the absolute value sign in real Jacobians. These trivial seeming ``sgn" factors will play an important role throughout our discussion in this paper and we must keep track of them everywhere they appear in real variable changes, for instance also as
\begin{equation}
\delta(a x) = \frac{1}{|a|} \delta(x)
\end{equation}
Continuing with the fourier integral, notice that on the support of two $\delta^2$ factors, the argument in the exponential is nicely
\begin{equation}
(a_1 \tilde \lambda_1 + a_2 \tilde \lambda_2)(\mu_3 + X \lambda_3) = a_1 (W_1 \cdot Z_3) + a_2 (W_2 \cdot Z_3)
\end{equation}
so we can pull this factor outside the $X$ integral and perform the remaining $X$ integral over the two $\delta^2$ factors that simply gives us $[1 2]^{-2}$. Putting everything together, we find
\begin{equation}
M^{++-}(W_1,W_2,Z_3) = {\rm sgn}([12]) \int \frac{d a_1}{a_1} e^{i a_1 (W_1 \cdot Z_3)} \int \frac{d a_2}{a_2} e^{i a_2(W_2 \cdot Z_3)}
\end{equation}
However, we have to make sense of the integral $\int \frac{da}{a} e^{i a x}$. This can be done by regulating $\frac{1}{a}$ as some linear combination of $\frac{1}{a + i \epsilon}$ and $\frac{1}{a - i \epsilon}$; our guiding principle for the correct $i \epsilon$ prescription here is to keep the little group properties of the amplitude manifest in twistor space; in order to ensure the amplitude has nice projective properties under rescaling by any real number, we must regulate with the principle value prescription $\frac{1}{a} \to \frac{1}{2}(\frac{1}{a + i \epsilon} + \frac{1}{a - i \epsilon})$, which sets
\begin{equation}
\int \frac{da}{a} e^{i a x} = {\rm sgn}(x)
\end{equation}
Note that this $i \epsilon$ prescription has nothing to do with
regulating propagators, after all we have encountered it here in
transforming the three-particle amplitude! It is necessary only to
keep the little group invariance manifest; we will later see however
that propagators must also be regulated with this principal value $i
\epsilon$ prescription.

The opposite $M^{--+}$ helicity configuration is naturally given in the $Z_1,Z_2,W_3$ basis as
\begin{equation}
M^{--+}(Z_1,Z_2,W_3) = {\rm sgn}(Z_1 I Z_2) {\rm sgn}(Z_1 \cdot W_3) {\rm sgn} (Z_2 \cdot W_3)
\end{equation}

The three-point amplitude in gravity can be determined by completely analogous arguments; if we use $W_1,W_2,Z_3$ for $M^{++-}$ then the amplitude
should have weight 2 under rescaling any of the variables; the sign functions simply get replaced by absolute values:
\begin{equation}
M^{++-}_{GR}(W_1,W_2,Z_3) = |W_1 I W_2| \, |W_1 \cdot Z_3| \, |W_2 \cdot Z_3|
\end{equation}
which can easily be verified by direct fourier transformation. Here $|x|$ is defined by the integral
$\int \frac{d a}{a^2} e^{i a x} = |x|$
with $\frac{1}{a^2}$ regulated by the principal value prescription.
The same object without the absolute value signs would have the correct little group
properties but is analogous to using ``1" for Yang-Mills theory, with singular support at zero momentum.

The extension to maximally supersymmetric amplitudes is straightforward. Just as in momentum space, the three-point amplitude is the sum of two terms,
\begin{equation}
M_{SYM} = M^{+}_{SYM} + M^{-}_{SYM}
\end{equation}
where $M^+$ contains the (++-) helicity amplitudes and $M^-$ the $(--+)$ helicity amplitudes. In twistor space, they are given by the obvious supersymmetrization of what we found above, replacing $W \to {\cal W}$, $Z \to {\cal Z}$:
\begin{eqnarray}
M^{+}_{SYM}({\cal W}_1,{\cal W}_2,{\cal Z}_3) &=& {\rm sgn} ({\cal W}_1 I {\cal W}_2)  \, {\rm sgn} ({\cal W}_1 \cdot {\cal Z}_3) \, {\rm sgn} ({\cal W}_2 \cdot {\cal Z}_3), \nonumber \\  M^{-}_{SYM}({\cal Z}_1,{\cal Z}_2,{\cal W}_3) &=&
{\rm sgn} ({\cal Z}_1 I {\cal Z}_2)  \, {\rm sgn} ({\cal Z}_1 \cdot {\cal W}_3) \, {\rm sgn} ({\cal Z}_2 \cdot {\cal W}_3)
\end{eqnarray}
Note that the expressions for $M^{+}$ and $M^{-}$ are not given in the same basis! Thus to explicitly write the amplitude in, say, ${\cal W}_1,{\cal W}_2, {\cal Z}_3$ basis, one of the terms appears naturally as a fourier-transform:
\begin{equation}
M_{SYM} = M^{+}({\cal W}_1,{\cal W}_2,{\cal Z}_3) + \int d^{4|4} {\cal Z}_1 d^{4|4} {\cal Z}_2 d^{4|4} {\cal W}_3
e^{i \sum_k {\cal Z}_k \cdot {\cal W}_k} M^{-}({\cal Z}_1,{\cal Z}_2,{\cal W}_3)
\end{equation}
The three particle amplitude for supergravity has exactly the same form with sgn($x$) replaced by $|x|$.

\subsection{The ``Link Representation" for Amplitudes}

Let us continue by looking at the 4-particle amplitude. These are of course no longer entirely determined by symmetries; for instance the cross-ratio $\frac{(Z_1 W_2)(Z_3 W_4)}{(Z_1 W_4) (Z_3 W_2)}$ is invariant under all rescalings and the amplitude could in principal be a general function of it. However as we will see the amplitudes continue to be remarkably simple. It will be convenient to look at the maximally supersymmetric four-particle amplitude in Yang-Mills, $M({\cal W}_1,{\cal Z}_2,{\cal W}_3,{\cal Z}_4)$. Fourier-transforming the known super-amplitude into twistor space can be done just as above, and we find
\begin{equation}
M({\cal W}_1,{\cal Z}_2,{\cal W}_3,{\cal Z}_4) = {\rm sgn}({\cal W}_1 \cdot {\cal Z}_2) {\rm sgn}({\cal Z}_2 \cdot {\cal W}_3) {\rm sgn}({\cal W}_3 \cdot {\cal Z}_4) {\rm sgn}({\cal Z}_4 \cdot {\cal W}_1)
\end{equation}
Note that every variable appears in a sign twice and therefore this expression has the correct weight.
From here we can read off various amplitudes in pure Yang-Mills; for instance putting all the $\eta,\tilde \eta \to 0$ we find
\begin{equation}
M^{+-+-}(W_1,Z_2,W_3,Z_4) = {\rm sgn}(W_1 \cdot Z_2) {\rm sgn}(Z_2 \cdot W_3) {\rm sgn}(W_3 \cdot Z_4) {\rm sgn}(Z_4 \cdot W_1)
\end{equation}
Setting $\eta_1,\tilde \eta_4 \to 0$ but integrating over $\eta_2, \tilde \eta_3$ yields the $(++--)$ amplitude; this is done conveniently by writing the sgn$(x)$ factors as $\int \frac{da}{a} e^{i a x}$ and we find
\begin{equation}
M^{++--}(W_1,Z_2,W_3,Z_4) = {\rm sgn}(W_1 \cdot Z_2) \delta^{\prime \prime \prime}(Z_2 \cdot W_3) {\rm sgn}(W_3 \cdot Z_4) {\rm sgn}(Z_4 \cdot W_1)
\end{equation}
where $\delta^{\prime \prime \prime}(x)$ arises from $\int \frac{da}{a} \times a^4 \times e^{i a x}$.

Returning to $M^{+-+-}$, it is natural to write it in the form
\begin{equation}
\label{simpleform} M^{+-+-}(W_i,Z_J) = \int \frac{dc_{12}}{c_{12}} \frac{d c_{14}}{c_{14}} \frac{d
c_{32}}{c_{32}} \frac{d c_{34}}{c_{34}} e^{ic_{iJ} W_i \cdot Z_J}
\end{equation}
A remarkable fact we will see in action later in this note, and elaborate on at greater length in \cite{HolS}, is that in an ambidextrous basis with sufficiently many (and at least two) $Z$'s and $W$'s, {\it any} amplitude can be written in this form:
\begin{equation}
M = \int dc_{iJ} \hat{M}(c_{iJ}; \tilde \lambda_i, \lambda_J) e^{i
c_{iJ} W_i \cdot Z_J}
\end{equation}
where the index $i$ runs over all particles labeled by $W$'s and $J$ over all particles labeled by $Z$'s.
This formula is telling us that all the dependence on the
$\mu's, \tilde \mu's$ is in the combination appearing in the exponentials!
This is extremely surprising, since a priori one might have expected that the amplitude can depend on complicated functions of many sorts of non-linear invariants like $(W_i W_j W_k W_l)$; the fact that the dependence on $W,Z$ is so strictly controlled is very striking. We call this the ``link representation" of the amplitude and the $c_{iJ}$ link variables.  For an $n$ point amplitude, we will see that the integrals over $c_{iJ}$ break up into pieces that each depend only on a small subset of all the possible link variables connecting $W_i,Z_J$; any piece will have only $2n - 4$ integrations. This justifies the nomenclature since the representation tells us about the way the $W_i,Z_J$ are linked up with each other.

If the twistor-space amplitude is given in the link representation, it is trivial to fourier-transform back to momentum space, since the integrals over the $\tilde \mu_i, \mu_J$ just give $\delta$ functions! We have
\begin{equation}
M(\lambda,\tilde \lambda) = \int d c_{iJ} \hat{M}(c_{iJ}; \tilde \lambda_i, \lambda_J) \delta^2 (\lambda_i - c_{i J}\lambda_J) \delta^2(\tilde \lambda_J + c_{iJ} \tilde \lambda_i)
\end{equation}
This is a remarkable formula. It has broken up the momentum
conservation $\delta$ function, which is quadratic in $\lambda,
\tilde \lambda$, into linear pieces: notice that
\begin{equation}
\lambda_i - c_{iJ} \lambda_J = 0, \, \tilde \lambda_J + c_{iJ} \tilde \lambda_i = 0 \implies \sum_i \lambda_i \tilde \lambda_i + \sum_J \lambda_J \tilde \lambda_J = 0
\end{equation}
Furthermore, as we mentioned, at $n$ points the amplitude breaks up
into pieces each of which only has $2n - 4$ link variables; thus,
there are always precisely enough $\delta^2$ functions to completely
determine the $c_{iJ}$'s by solving a series of linear equations,
leaving us with the momentum-conserving $\delta$ function.
Therefore, getting the momentum space amplitude from the link
representation involves no integrations whatsoever, but merely
solving a set of linear equations to determine the $c_{iJ}$
\footnote{This is very reminiscent of the RSV formula for the tree
S-Matrix in ${\cal N} = 4$ SYM \cite{RSV}. Their expression follows
from transforming back to momentum space the connected prescription
for computing amplitudes in Witten's twistor string theory. A very
important difference is that they had to solve highly non-linear
equations, while our amibidextrous formulation reduces to solving
linear equations.}.

Let us see how this works for $M^{+-+-}$, where the link representation is
\begin{equation}
M^{+-+-} = \int d c_{iJ} \frac{1}{c_{12}c_{14} c_{32} c_{34}} \delta^2(\lambda_i - c_{iJ} \lambda_J) \delta^2(\tilde \lambda_J + c_{iJ} \tilde \lambda_i)
\end{equation}
We see that the $c_{iJ}$ can be explicitly solved for, though there are a number of different forms the solution can take that are all equivalent on the support of the momentum-conserving delta function. For instance, just from the equations for $\lambda_1 = c_{12} \lambda_2 + c_{14} \lambda_4$, and $\lambda_3 = c_{32} \lambda_2 + c_{34} \lambda_4$ we can determine
\begin{equation}
c_{i2} = \frac{\langle i 4 \rangle}{\langle 2 4 \rangle}, \, c_{i4} = \frac{\langle i 2 \rangle}{\langle 4 2 \rangle}
\end{equation}
and it is trivial to see that the Jacobian in replacing with the two $\delta^2$ factors with the single $\delta$'s fixing the $c_{iJ}$ precisely cancels against the one that converts the remaining two $\delta^2$ factors into a single momentum conserving $\delta$ function. We thus find $M^{+-+-} = {\cal M}^{+-+-} \delta^4(\sum_k p_k)$ with
\begin{equation}
{\cal M}^{+-+-} = \frac{1}{c_{12} c_{32} c_{34} c_{14}} = \frac{\langle 2 4 \rangle^4}{\langle 1 2 \rangle \langle 2 3 \rangle \langle 3 4 \rangle \langle 4 1 \rangle}
\end{equation}
recovering the familiar MHV form of the
amplitude~\cite{ParkeTaylor}. We could have also chosen to solve for
the $c_{iJ}$ from the $\tilde \lambda$ equations, and that would
have given us a different equation with the $\overline{{\rm MHV}}$
form of the amplitude, or we could have solved for $c_{12},c_{14}$
from the $\lambda_1$ equation and the $c_{32},c_{34}$ from the
$\tilde \lambda_2$ equation, giving us a mixed form of the 4
particle amplitude. These are all different familiar representations
of the 4 particle amplitude, which are equal to each other due to
momentum conservation. This highlights that in a sense, the link
form of the amplitude describes the amplitude in the most invariant
way, and only the insistence to factor out the momentum conserving
delta function introduces asymmetries in how the amplitude is
written. It is very pleasing that the form of the amplitude in
twistor space immediately leads to this most invariant form of the
amplitude back in momentum space!

Let us make another  comment about the link representation.
Suppose we are given an amplitude with some number of $Z'$s and
$W'$s in the link representation, and suppose that some pair
$W_{i_*}, Z_{J_*}$ are indeed linked. Then we can decide to change
$i_*$ to the $Z$ representation and $J_*$ to the $W$
representation; this obviously keeps the net number of $Z$'s and
$W$'s unchanged, and in this new basis the amplitude will also
have a link representation. We can illustrate this for the full
super-amplitude, where we can e.g. switch from $M({\cal W}_1,{\cal
Z}_2,{\cal W}_3,{\cal Z}_4)$ to $M({\cal W}_1,{\cal W}_2, {\cal
Z}_3,{\cal Z}_4)$. A quick computation gives
\begin{equation}
\label{wwzz}
M({\cal W}_1, {\cal W}_2,{\cal Z}_3, {\cal Z}_4) = \int dc_{iJ} \frac{1}{c_{13} c_{24} (c_{13} c_{24} - c_{14} c_{23})} e^{i c_{iJ} {\cal W}_i \cdot {\cal Z}_J}
\end{equation}
Sending the $\eta, \tilde \eta \to 0$ gives a link
representation of the $M^{++--}$ amplitude in pure Yang-Mills.

Finally, as a sample gravitational amplitude, we write a link
representation of the 4pt amplitude for ${\cal N} = 8$ SUGRA, in
the ${\cal W}_1,{\cal W}_2,{\cal Z}_3,{\cal Z}_4$ basis, which is
the beautifully symmetrical object
\begin{equation}
M^{{\rm SUGRA}}({\cal W}_1,{\cal W}_2,{\cal Z}_3,{\cal Z}_4) = \int d c_{iJ} \frac{[12]\langle 3 4 \rangle}{c_{13} c_{14} c_{23} c_{24} (c_{13} c_{24} - c_{14} c_{23})} e^{i c_{iJ} {\cal W}_i \cdot {\cal Z}_J}
\end{equation}

\section{BCFW in Twistor Space}

We now show that with (2,2) signature, the BCFW recursion relations find their most natural home in twistor space. Indeed, even if we had never heard of twistor space, the most natural formulation of the BCFW formula in (2,2) signature would force us to discover it!

\subsection{The Recursion Relation in Twistor Space}

Before plunging into the derivation, we jump ahead to giving the
final result, in order to emphasize that it is essentially the only
possible natural expression we could have written down in twistor
space. This will also give us the opportunity to introduce some of
the objects that will appear so that we can better understand them
when they arise in the derivation. We start with pure Yang-Mills,
even though the formulas are most compact and beautiful for the
maximally supersymmetric case. For pure Yang-Mills, there is a BCFW
formula for deforming particles $i$ and $j$ as long as the
helicities $(h_i,h_j) \neq (-,+)$. In the case where the helicity is
$(+,-)$ and we work in the $W_i,Z_j$ basis, the BCFW formula is
\begin{eqnarray}
\label{bcfwYM}
&M(W_i,Z_j) = \sum_{L,R} \int \left[D^3 Z_P D^3 W_P \right]_{W_i,Z_j} &\nonumber \\ &\left[M_L(W_i;Z_P,+) M_R(Z_j; W_P,-) + M_L(W_i;W_P,-) M_R(Z_j;Z_P,+) \right]&
\end{eqnarray}
where we have suppressed the dependence on the $Z$'s and $W$'s which
label the remaining external particles. Some comments are in order.
Most strikingly, note that unlike the BCFW formula in momentum
space, {\it there is no deformation of the twistor variables
appearing in the amplitudes}, that is, the particles $i,j$ are
represented by the same twistors $W_i,Z_j$ on the left and right
hand side of the equations. Whereas in the usual BCFW formula we
have the internal propagator, the internal particle is now labeled
by $W_P,Z_P$, which are integrated over. Here the subscript ``$P$"
refers to the fact that these are projective variables on
$\mathbb{RP}^3$, which was to be expected. The symbol $[D^3 W_P D^3
Z_P]_{W,Z}$ denotes a projective measure:
\begin{eqnarray}
\left[D^3 W_P D^3 Z_P\right]_{W_i,Z_j} = \, D^3W_P D^3 Z_P \, \nonumber \\ \times {\rm sgn} (W_i \cdot Z_j) \delta^{\prime \prime \prime}(W_P \cdot Z_P) {\rm sgn}(W_P I W_i) {\rm sgn}(Z_P I Z_j)
\end{eqnarray}

The measures $D^3 W_P$, $D^3 Z_P$ are in turn the natural projective measure on $\mathbb{RP}^3$, which we can define more generally for any $\mathbb{RP}^{n-1}$. Consider co-ordinates $X^A$ in $\mathbb{R}^n$, we can define co-ordinates on $\mathbb{RP}^{n - 1}$ via $X^A = u X_P^A$ where $X^A_P = (1, x_1, \cdots, x_{n-1})$. Then
\begin{equation}
d^n X = du |u|^{n - 1} dx_1 \cdots dx_{n-1} \to  du |u|^{n-1} \epsilon_{A B_1 \cdots B_{n-1}} X^A_P \wedge d X_P^{B_1} \cdots \wedge dX_P^{B_{n-1}} \nonumber \equiv  d u |u|^{n-1} D^{n-1} X_P \nonumber
\end{equation}
We can in fact see that $[D^3 W_P D^3 Z_P]_{W_i,Z_j}$ is essentially the only natural measure we can use to projectively integrate functions $F(W,Z)$ of weight zero over twistor space. Note that the factor $\delta^{\prime \prime \prime}(W_P\cdot Z_P)$ {\it almost} has weight -4,  under rescaling e.g. $W_P \to \rho W_P$ it changes as $\rho^{-4}$ sgn $\rho$ which, were it not for the sgn $\rho$ factor, would cancel the weight of the projective measure $D^3 W_P$. To cancel the extra factor of sgn$\rho$ and  have a well-defined measure against which we can integrate functions of weight zero, there must be an additional factor involving sgn$(W_P \cdot Z_{ref})$ and sgn $(Z_P \cdot W_{ref})$ for some reference $Z_{ref}$ and $W_{ref}$. The only natural reference objects available are $W_i,Z_j$, so we can have e.g. either $Z_{ref} = Z_j$ or $Z_{ref} = I W_i$. As our derivation will show, the BCFW formula makes the latter choice. Note that with these additional sgn factors, under rescaling $W_i \to \rho_i W_i$, $Z_j \to \kappa_j Z_j$, the measure now picks up a factor sgn($\rho_i \kappa_j$). In order to cancel this factor and be left with a function with zero projective weights under rescaling $W_i,Z_j$, we should multiply by an additional factor of sgn$(W_i \cdot Z_j)$. Thus, we see the integral over $\left[D^3 W_P D^3 Z_P \right]_{W_i,Z_j}$ is essentially the unique way of integrating the nice object of weight zero $M_L M_R$ over twistor space, to yield another function of weight zero. This then yields a natural way of building higher point amplitudes from lower ones. One can show that using the second choice for the measure to define higher-point amplitudes via a BCFW-type formula yields momentum space amplitudes with more singular support than just the momentum-conserving delta function, and we have not yet found a nice physical interpretation for it.

Note also the appearance of the factor $\delta^{\prime \prime \prime}(W_P \cdot Z_P)$. This bears a striking resemblance to Witten's 1978 formulation of Yang-Mills theory in twistor space~\cite{Witten78}, which demanded the existence of a holomorphic bundle on the third neighborhood of the ``quadric" $W \cdot Z = 0$!

With maximal SUSY, the BCFW formula in twistor space is even more compact and elegant:
\begin{equation}
\label{bcfwSYM}
M({\cal W}_i,{\cal Z}_j) = \sum_{L,R} \int \left[D^{3|4} {\cal W}_P D^{3|4} {\cal Z}_P\right]_{{\cal W}_i,{\cal Z}_j} M_L({\cal W}_i, {\cal Z}_P) M_R({\cal Z}_j; {\cal W}_P)
\end{equation}
where
\begin{eqnarray}
\left[D^{3|4} {\cal W}_P D^{3|4} {\cal Z}_P\right]_{{\cal W}_i,{\cal Z}_j} = \, D^{3|4}{\cal W}_P D^{3|4}{\cal Z}_P \, \nonumber \\ \times {\rm sgn}({\cal W}_i \cdot {\cal Z}_j) {\rm sgn}({\cal W}_P \cdot {\cal Z}_P) {\rm sgn} ({\cal W}_P I {\cal W}_i) {\rm sgn}({\cal Z}_P I {\cal Z}_j)
\end{eqnarray}
In comparing to pure-Yang-Mills, in addition to the straightforward changes of $(W,Z) \to ({\cal W},{\cal Z})$, in the measure the $\delta^{\prime \prime \prime}(W \cdot Z)$ has been replaced by sgn$({\cal W} \cdot {\cal Z})$; we can see that integrating over the $\tilde \eta, \eta$ takes four derivatives of this object and converts it into $\delta^{\prime \prime \prime}(W_P \cdot Z_P)$. Put another way, up to sgn factors the extra Grassmann Jacobian in rescaling e.g. ${\cal W} \to \lambda {\cal W}$ already gives $D^{3|4}{\cal W}$,$D^{3|4}{\cal Z}$ the correct weight zero; the sgn factors ensure that this works out correctly for rescalings by any real number. Note that had we used a formalism with manifest ${\cal N} = 3$ SUSY, we would find a factor $\delta({\cal W}_P \cdot {\cal Z}_P)$ instead; this again resonates with Witten's 1978 work, since with ${\cal N} = 3$ SUSY he found that the SYM equations of motion could be determined directly on the super-quadric ${\cal W}_P \cdot {\cal Z}_P = 0$, without the need to go to its third neighborhood.

Note also that for ${\cal N} =4$ SYM, there is no helicity sum on the internal line, just as in momentum space. This is one of the beautiful  and unique features of maximally supersymmetric theories: the SUSY multiplet is CPT self-conjugate, and unifies positive and negative helicities. As a consequence of this freedom, there is a second representation of the BCFW formula with ${\cal W}_P,{\cal Z}_P$ swapped in the projective integral.

While the presence of these projective integrals is natural, one may be put off at the prospect of having to do non-linear integrals to get amplitudes. In fact these projective integrals can immediately be ``de-projectivized" into integrals over a full $d^4 W d^4 Z$. There are many ways of doing this, and different choices can be useful in different situations, but a canonical way of doing it is as follows. Consider any function $F(W,Z)$ of weight zero under rescaling $W,Z$, and look at the projective integral
\begin{eqnarray}
I &=& \int \left[D^3 W_P D^3 Z_P\right]_{W_i,Z_j} F(W_P,Z_P) \nonumber \\ &=& {\rm sgn}(W_i \cdot Z_j) \int D^3W_P D^3 Z_P \delta^{\prime \prime \prime}(W_P \cdot Z_P) {\rm sgn}(W_P I W_i) {\rm sgn}(Z_P I Z_j) F(W_P,Z_P)
\end{eqnarray}
Let us express sgn$(W_P I W_i) = \int \frac{du}{u} e^{i u (W_P I W_i)}$ and sgn$(Z_P I Z_j) = \int \frac{dv}{v} e^{i v (Z_P I Z_j)}$. We would like to write this integral as one over $W = u W_P$ and $Z = v Z_P$. If we note that $\delta^{\prime \prime \prime}(W_P \cdot Z_P) = u^4 \, {\rm sgn} u \, \, v^4 {\rm sgn} v \, \delta^{\prime \prime \prime}(u W_P \cdot v Z_P)$, and $F(W_P,Z_P)$ = $F(u W_P,v Z_P)$, we are left with an integral depending only on $W = u W_P$ and $Z = u Z_P$, with measure $D^3 W_P du |u|^3 \times D^3 Z_P dv |v|^3 = d^4 W d^4 Z$. We have thus deprojectivized the integral as
\begin{equation}
I = {\rm sgn}(W_i \cdot Z_j) \int d^4 W d^4 Z \delta^{\prime \prime \prime}(W \cdot Z) e^{i W I W_i} e^{i Z I Z_j} F(W,Z)
\end{equation}
The supersymmetric integrals can be similarly de-projectivized, replacing $D^{3|4} {\cal W}_P D^{3|4} {\cal Z}_P \to d^{4|4} {\cal W} d^{4|4} {\cal Z}$ and the sgn${\cal W}_P I {\cal W}_i$ sgn${\cal Z}_P I {\cal Z}_j$ factor with $e^{i {\cal W} I {\cal W}_i} \, e^{i {\cal Z} I {\cal Z}_j}$. Indeed, these expressions can be thought of as providing an alternate definition of our projective integrals.

We reassure the reader still daunted at the prospect of performing explicit integrals over twistor space that, even though we have taken some pains to talk about these projective integrals properly, we will never have to compute any non-trivial integrals! In the next section we will instead introduce a simple diagrammatic formalism for the BCFW recursion relations in twistor space, which will allow us to graphically manipulate these objects in an efficient way.

The recursion relations for gravity and supergravity take exactly
the same form; since we are integrating amplitudes of weight $2$
rather than $0$, only the measure is trivially altered:
\begin{eqnarray}
\left[D^3 W_P D^3 Z_P\right]_{W_i,Z_j} = D^3W_P D^3 Z_P \nonumber \\ \times {\rm sgn} (W_i \cdot Z_j) \, \delta^{\prime \prime \prime \prime \prime}(W_P \cdot Z_P) {\rm sgn}(W_P I W_i) \, {\rm sgn}(Z_P I Z_j) \nonumber
\end{eqnarray}
and
\begin{eqnarray}
\left[D^{3|8} {\cal W}_P D^{3|8} {\cal Z}_P\right]_{{\cal W}_i,{\cal Z}_j} = \, D^{3|8}{\cal W}_P D^{3|8}{\cal Z}_P \nonumber \\  \times {\rm sgn}({\cal W}_i \cdot {\cal Z}_j) \,  ({\cal W}_P \cdot {\cal Z}_P)^2 {\rm sgn}({\cal W}_P \cdot {\cal Z}_P) \, {\rm sgn} ({\cal W}_P I {\cal W}_i) {\rm sgn}({\cal Z}_P I {\cal Z}_j)
\end{eqnarray}

\subsection{Into Twistor Space Via Fully On-Shell BCFW}
We now proceed to show that the most natural and maximally on-shell formulation of the BCFW recursion relation in $(2,2)$ signature begs to be fourier transformed into twistor space, leading to eqn (\ref{bcfwYM}). Deforming particles $(i,j)$ with helicities $(+,-)$, the BCFW recursion relation is
\begin{equation}
\label{bcfw} {\cal M} = \sum_{L,h} {\cal M}_L(p_i(\tau_P), \{-P_L(\tau_P), h\}, L) \frac{1}{P_L^2} {\cal
M}_R(p_j(\tau_P), \{P_R(\tau_P), -h\}, R).
\end{equation}
where
\begin{equation}
\lambda_i(\tau) = \lambda_i + \tau \lambda_j, \, \, \tilde
\lambda_j(\tau) = \tilde \lambda_j - \tau \tilde \lambda_i
\end{equation}
and
\begin{equation}
\tau_P = - \frac{P_L^2}{[i|P_L|j \rangle}.
\end{equation}
We use ``$\tau$" rather than the more customary ``$z$" in these
expressions, because we want to emphasize that in split signature,
all the variables and in particular the $\tau_P$ are real, which
will be crucial for our entire discussion. Almost all elements
entering in the form (\ref{bcfw}) are on-shell, except for the
explicit propagator $1/P_L^2$ which is off-shell. There is an even
more natural way of writing the recursion relation in a form that is
manifestly on-shell.

Let us consider the physical amplitude $M$ including the momentum-conserving
delta function as
\begin{equation}
M = \delta^{4}(\sum_{k=1}^n p_k) {\cal M},
\end{equation}
then (\ref{bcfw}) is equivalent to
\begin{eqnarray}
\label{fami}
 M = \sum_{L,h} {\rm sgn}(-[i|P_L|j\rangle )\int_{-\infty}^{\infty}\frac{d\tau}{\tau}\int d^4P \delta(P^2) \nonumber \\
M_L(p_i(\tau), \{-P, h\}, L) \, M_R(p_j(\tau), \{P,
-h\}, R).
\end{eqnarray}
This can be easily checked by using the delta function in $M_L$ to
perform the $d^4P$ integral and $\delta(P^2)$, which becomes
$\delta(\tau\langle j|P_L|i] - P^2_L)$, to perform the $\tau$
integral. Note that there is strictly speaking no reason to regulate
this $\tau$ integral in any way, since the delta functions fix $\tau
= \tau_P$ which does not vanish for generic external momenta.
However, when we later fourier transform, reverse orders of
integration and so on, we will have to be more careful. By now we are accustomed to seeing a factor like sgn$[i|P_L|j \rangle$, which is again there because, for real variables, $\delta (a x) = \frac{1}{|a|} \delta(x) = \frac{{\rm sgn} a}{a} \delta(x)$.
Note that $\delta(\tau\langle j|P_L|i] - P^2_L)$ always has support on the integration again since all our variables are real in split signature.

Since the
momentum conserving delta function forces $P = - P_L + \tau_P |j\rangle[i|$, we can replace the sgn$[i|P_L|j\rangle$ with sgn$[i|P|j\rangle$ underneath the $P$ integral, obtaining
\begin{eqnarray}
\label{fami2} M = \sum_{L,h}
\int_{-\infty}^{\infty}\frac{d\tau}{\tau}\int d^4P \delta(P^2) {\rm
sgn}([i|P|j\rangle) \nonumber \\  M_L(\{p_i(\tau), h_1\}, \{-P, h\}, L)
M_R(\{p_j(\tau), h_2\}, \{P, -h\}, R).
\end{eqnarray}

Now, the BCFW
deformation $\lambda_i(\tau) = \lambda_i + \tau \lambda_j$, $\tilde
\lambda_j = \tilde \lambda_j - \tau \tilde \lambda_i$, is just a
translation in $\lambda_i$, and a separate translation in $\tilde
\lambda_j$, while the other variables are unchanged. As usual when
we have translations, it is natural to fourier transform to the
conjugate momentum basis in order to diagonalize the translations as
multiplication by a phase. This is how the BCFW formula forces us to discover twistor space! Fourier transforming $\int d^2 \lambda_i e^{i \tilde \mu_i \lambda_i} \int d^2 \tilde \lambda_j e^{i \mu_j \tilde \lambda_j}$, all
the $\tau$ dependence in the product $M_L M_R$ is extracted as a phase factor
\begin{equation}
e^{i \tau \tilde \mu_i \lambda_j} e^{i (-\tau) \mu_j \tilde \lambda_i} = e^{i \tau W_i \cdot Z_j}
\end{equation}
We can then isolate the $\tau$ integral, yielding
\begin{equation}
\int_{-\infty}^{\infty} \frac{d \tau}{\tau} e^{i \tau (W_i \cdot Z_j)}
\end{equation}
We last encountered such an object in the explicit fourier transform of the three-particle amplitude into twistor space; we emphasized there that regulating the integral with the principal value prescription had nothing to do with the usual propagator $i \epsilon$ choice, and instead was dicated by getting the correct projective property for the twistor space amplitude. In the present case, demanding the correct projective property once again forces
\begin{equation}
\int_{-\infty}^{\infty} \frac{d \tau}{\tau} e^{i \tau (W_i \cdot Z_j)}
 \to {\rm sgn}(W_i \cdot Z_j)
\end{equation}
but this time, this choice does naturally correspond to choosing the principal value prescription to regulate the propagators at tree level. To see explicitly what the principal value prescription is buying us, suppose we instead use the usual Feynman $i \epsilon$ prescription $1/P_L^2 \to 1/(P_L^2 + i \epsilon)$ in equation (\ref{bcfw}). Then after doing the $P$ integral we would be left with $1/(\tau + i \epsilon {\rm sgn}\langle j|P_L|i])$, i.e. the $\tau$ integral would be regulated in a different way for different terms in the BCFW sum! This would make it impossible to bring the sgn factor inside the integral and continue as we did above. Instead, using the principal value prescription (symmetrizing with respect to the sign of $\epsilon$) corresponds to using the same principal value prescription for $\tau$, and allows to nicely transform to twistor space. We will give a more complete discussion of this issue in our discussion of loop amplitudes.

We have therefore arrived at the following form of the recursion
relation, with the external particles $i,j$ transformed to twistor
space in the $W_i,Z_j$ representation:
\begin{eqnarray}
\label{fami2} M(W_i,Z_j) = {\rm sgn}(W_i \cdot Z_j) \sum_{L,h}
\int d^4P \delta(P^2) {\rm sgn}([i|P|j\rangle) \nonumber \\
M_L(W_i, \{-P, h\}, L) M_R(Z_j, \{P, -h\}, R).
\end{eqnarray}
It is natural to continue the trend of casting everything in the most on-shell form possible by re-writing the integral over the phase space factor $d^4 P \delta(P^2)$.
This is familiar from the usual Minksowski signature, where $d^4 P
\theta(P^0) \delta(P^2)$ is written as a contour integral on
$\mathbb{R}^+\times \mathbb{CP}^1\times\mathbb{CP}^1$ with contour
the diagonal $\mathbb{CP}^1$~\cite{CSW}. We are after the analogous formula
with $\mathbb{RP}^1$'s.

Beginning with $\delta(P^2) = \delta(P_{1\dot 1}P_{2\dot 2}-P_{1\dot
2}P_{2\dot 1})$, we can e.g. integrate over $P_{1\dot 1}$ in order
to write the measure as
\begin{equation}
\label{measure} \frac{dP_{2\dot 2}dP_{1\dot 2}dP_{2\dot
1}}{|P_{2\dot 2}|}
\end{equation}
We can parametrize the on-shell momentum that appear here as $P_{a
\dot{a}} = t \lambda_a \tilde \lambda_{\dot{a}}$, where each spinor
is to be thought of as a homogeneous co-ordinate on one
$\mathbb{RP}^1$. More concretely, we can write $P_{1\dot 2} =t l,
P_{2\dot 1}=t \tilde l$ and $P_{2\dot 2}=t$, where $\lambda=(l,1)$
and $\tilde\lambda=(\tilde l, 1)$ are inhomogeneous co-ordinates on
$\mathbb{RP}^1$. One finds that (\ref{measure}) becomes
\begin{equation}
\label{kama} \int d^4P \delta(P^2) = \int_{-\infty}^{\infty} |t| dt
\int  D \lambda_P D \tilde \lambda_P
\end{equation}
where $D \lambda_P, D \tilde \lambda_P$ are the projective measures we have previously defined.
There are  two differences from the familiar form of this measure in
the ordinary Minkowski case. First, the $D\lambda_P D \tilde \lambda_P$ integral is no
longer a contour integral but an integral over the whole
real space,  and second, the integral over $t$ is over all real $t$ and not just $t>0$.
The product $M_L M_R$ is then a function of $t, \lambda_P, \tilde \lambda_P$ of the form
$M_L(t \lambda_P, \tilde \lambda_P, h) M_R(t \lambda_P, - \tilde \lambda_P, -h) = M_L(\lambda_P, - t \tilde \lambda_P, h) M_R(\lambda_P,t \tilde \lambda_P, -h)$. Using the little group, we can put these in a more symmetrical form:
for $h=+$ we write $t^2 M_L(t \lambda_P, \tilde \lambda_P, +) M_R(\lambda_P, - t \tilde \lambda_P, -)$, while for $h=-$ we write $t^2 M_L(\lambda_P, -\tilde \lambda_P, -) M_R (t \lambda_P, \tilde \lambda_P, +)$.
We then complete the transition to twistor space by writing $M_L$ and $M_R$ as the inverse fourier-transform from twistor space. For $h = +$ we write
\begin{eqnarray}
M_L(t \lambda_P,\tilde \lambda_P,+) = \int d^2 \tilde \mu e^{-i \tilde \mu t \lambda_P} M(W_P,+) \nonumber \\  M_R(\lambda_P,-t \tilde \lambda_P,-) = \int d^2 \mu e^{- i \mu (-t \tilde \lambda_P)} M(Z_P,-)
\end{eqnarray}
We have now assembled all the pieces. Clearly the $M_L M_R$ product becomes
\begin{equation}
t^2 M_L(W_P,+) M_R(Z_P,-) e^{i t Z_P \cdot W_P}
\end{equation}
the integration measure is
\begin{equation}
dt |t| D \lambda_P D \tilde \lambda_P d^2 \tilde \mu d^2 \mu = dt |t| D^3 W_P D^3 Z_P
\end{equation}
The sgn($[i| P |j \rangle$) factor becomes
\begin{equation}
{\rm sgn}[i|P|j\rangle = {\rm sgn}(t) {\rm sgn}([i \tilde \lambda_P]) {\rm sgn}(\langle \lambda_P j \rangle) = {\rm sgn}(t) {\rm sgn}
(W_P I W_i) {\rm sgn}(Z_P I Z_j)
\end{equation}
Putting everything together, we get the twistor space form of the BCFW recursion relation given in equation (\ref{bcfwYM}); the $\delta^{\prime \prime \prime}(W_P \cdot Z_P)$ simply represents
\begin{equation}
\delta^{\prime \prime \prime}(Z_P \cdot W_P) = \int dt t^3 e^{i t Z_P \cdot W_P}.
\end{equation}
Completely analogous steps lead to recursion relation for ${\cal N} = 4$ SYM, as well as gravity and ${\cal N} = 8$ supergravity. Here we begin with the supersymmetric form of the BCFW recursion relation \cite{simplest,Brand,moresusyBCFW}. By using the $\eta$,$\tilde \eta$ variables, any pair of particles can be deformed. What makes this possible is an associated deformation of the Grassmann parameters, which is the supersymmteric analog of the BCFW deformation on the momenta. Using the $\tilde \eta_i$ and $\eta_j$ representations and following the same steps above, the deformed $\eta$'s precisely have structure to turn the sgn$(W_i \cdot Z_j)$ factor into sgn$({\cal W}_i \cdot {\cal Z}_j)$. Expressing the sum over the internal particle states as $\int d^{\cal N} \eta d^{\cal \tilde \eta} e^{ \eta \cdot \tilde \eta} M_{L}(\tilde \eta) M_R (\eta)$ then turns the e.g. $\delta^{\prime \prime \prime}(W_P \cdot Z_P)$ factor in Yang-Mills into sgn$({\cal W}_P \cdot {\cal Z}_P)$ for SYM, and $\delta^{\prime \prime \prime \prime \prime}(W_P \cdot Z_P)$ into $({\cal W}_P \cdot {\cal Z}_P)^2$ sgn $({\cal W}_P \cdot {\cal Z}_P)$ for SUGRA.

\section{BCFW and ``Hodges Diagrams"}

There is a very natural diagrammatic representation of amplitudes in
twistor space, that greatly simplifies the BCFW computation of scattering
amplitudes. We call these diagrams ``Hodges diagrams" after Andrew
Hodges, who introduced very similar diagrams in the course of his work
on twistor diagrams~\cite{Hodges}. Indeed our diagrams are decorated with extra
features absent in Hodges' diagrams, which serve to make his diagrams
perfectly well-defined!

It is possible to present Hodges diagrams for both the
non-supersymmetric and maximally supersymmetric theories. For
introductory purposes it would probably be a little simpler to first see the
nuts and bolts of the diagrams in action in the non-supersymmetric
setting, which also gives an appreciation for the power of the
supersymmetric formalism for unifying many non-supersymmetric
amplitudes into a single supersymmetric object (in a way that goes
well beyond the familiar Ward identities). We will defer this more
complete discussion to \cite{HolS}, however, and immediately present
the maximally supersymmetric version of the diagrams,
which in fact look simpler than their non-supersymmetric cousins.

\subsection{Notation}

Let us begin with some basic notation. We will denote twistor
variables ${\cal Z}$ with a black dot and dual twistor variables
${\cal W}$ with a white dot. The commonly encountered ``sgn" factors
in the amplitudes will be denoted by a line connecting dots: a sgn$
{\cal W} \cdot {\cal Z}$ factor with a solid black line connecting
the corresponding white and black dots, and sgn${\cal Z}_1 I {\cal
Z}_2$ or sgn${\cal W}_1 I {\cal W}_2$ with a dashed black line
connecting black to black or white to white dots. A squiggly line
between ${\cal Z},{\cal W}$ will denote the factor $e^{i {\cal Z}
\cdot {\cal W}}$. This notation is summarized below:

\be \label{Notation}
\includegraphics[scale=0.5]{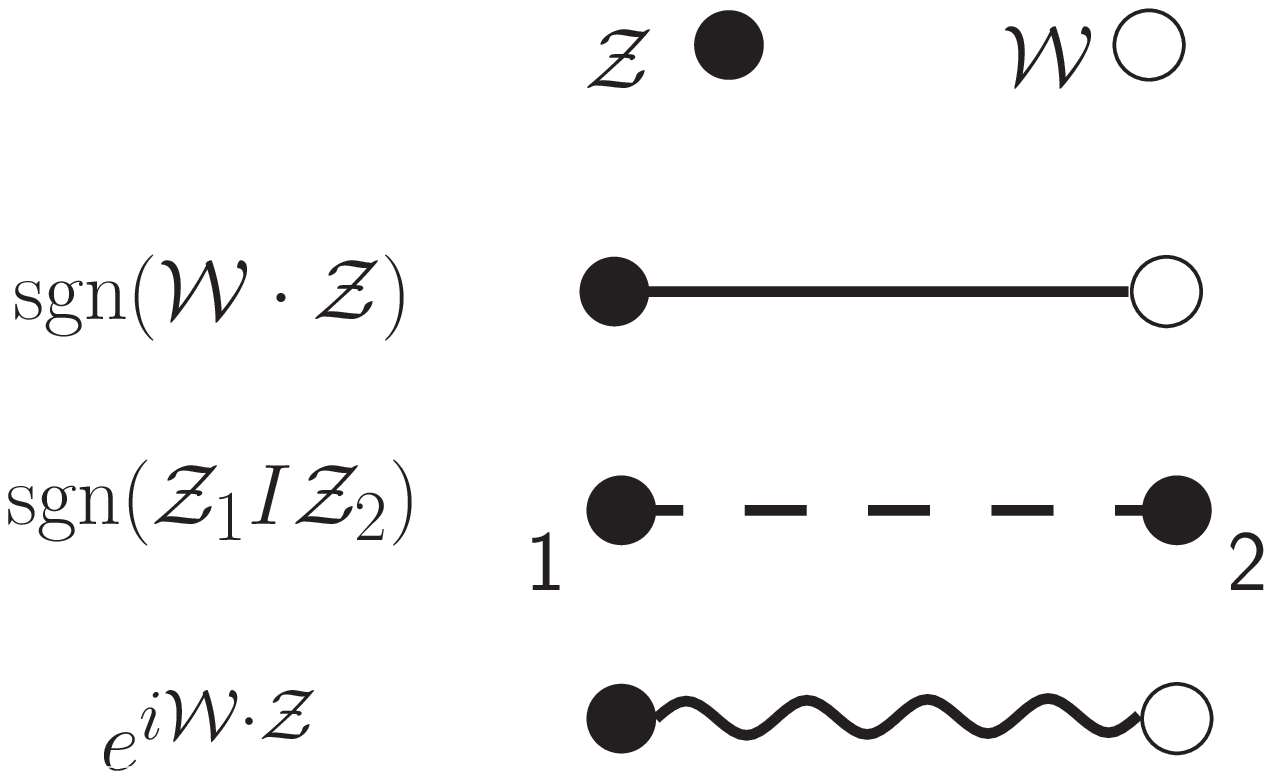} \nonumber
\ee

\vspace{0.3cm}

A general amplitude can be written in any basis we like, with ${\cal W}$'s
labeling some particles and ${\cal Z}$'s labeling others. We can freely go back
and forth between different representations by twistor
transformation e.g. $\int d^{4|{\cal N}} {\cal W} e^{i {\cal W} \cdot {\cal Z}}$, which can be graphically denoted by the addition of
squiggly lines as below:
\vspace{0.25cm} \be
\includegraphics[scale=0.6]{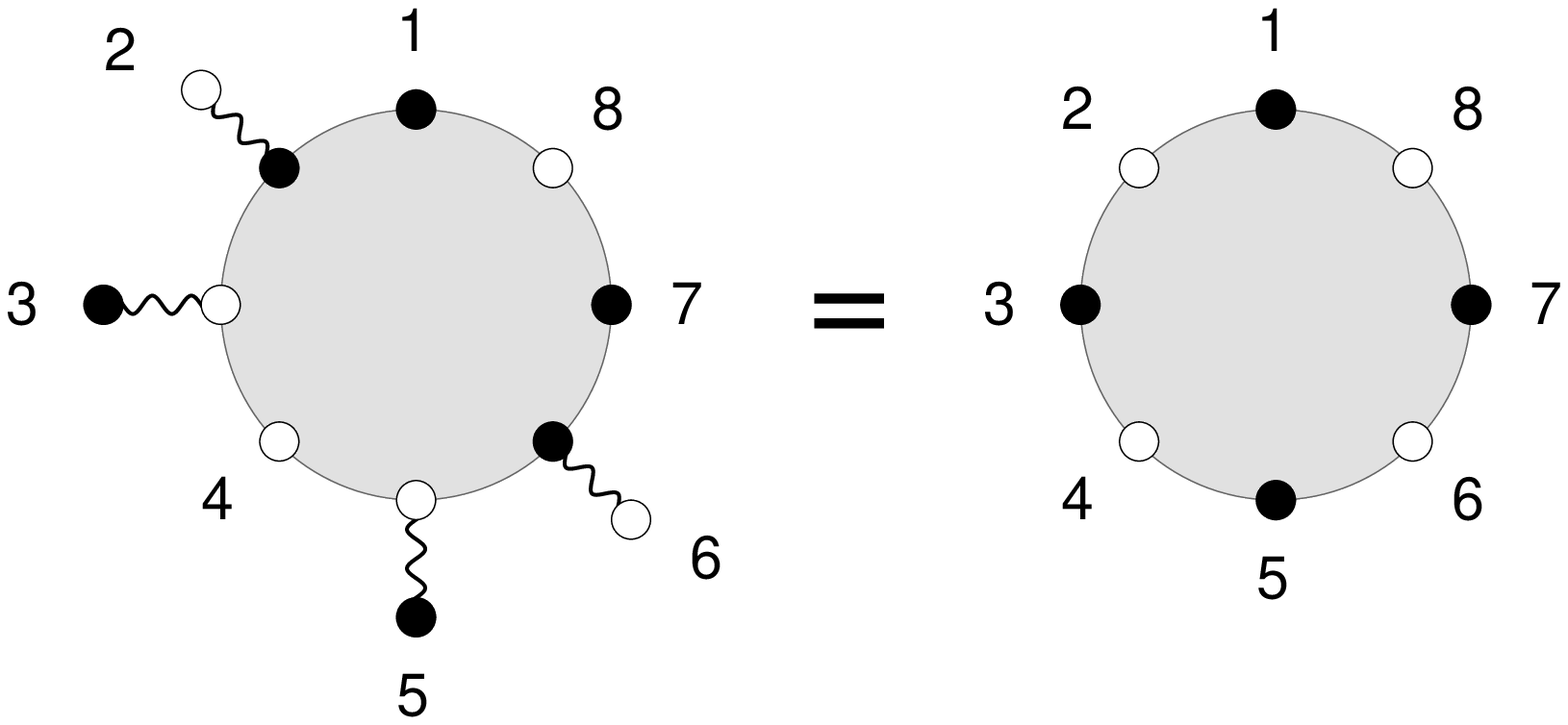} \nonumber
\ee
where unless otherwise specified, unlabeled white and black dots are to be integrated over
$d^{4|{\cal N}} {\cal W}$, $d^{4|{\cal N}} {\cal Z}$.

With this notation, the three-point functions $M_3^+$ and $M_3^-$ are represented by the Hodges diagrams shown below

\be
\includegraphics[scale=0.6]{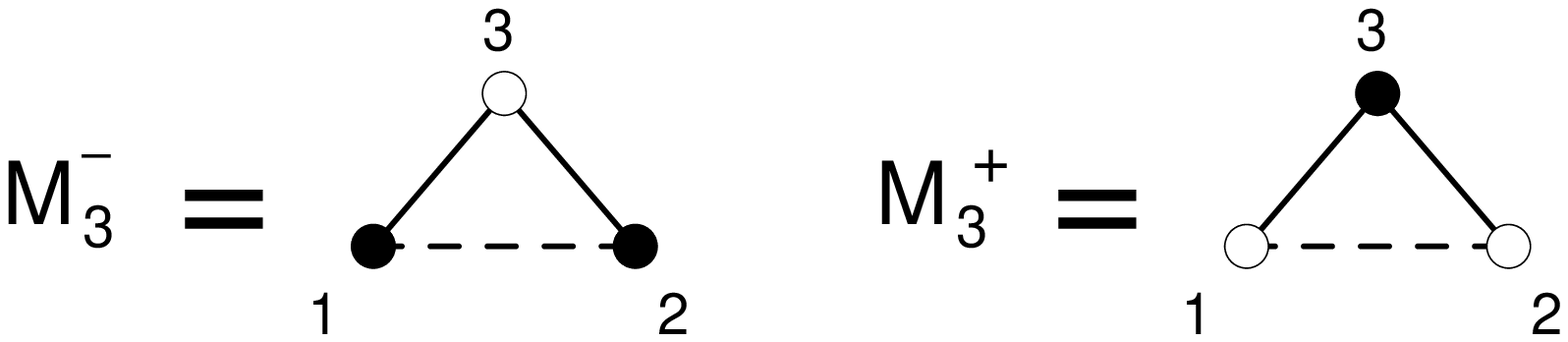} \nonumber
\ee

The Hodges diagram for the four-particle amplitude is also very simple

\be
\includegraphics[scale=0.4]{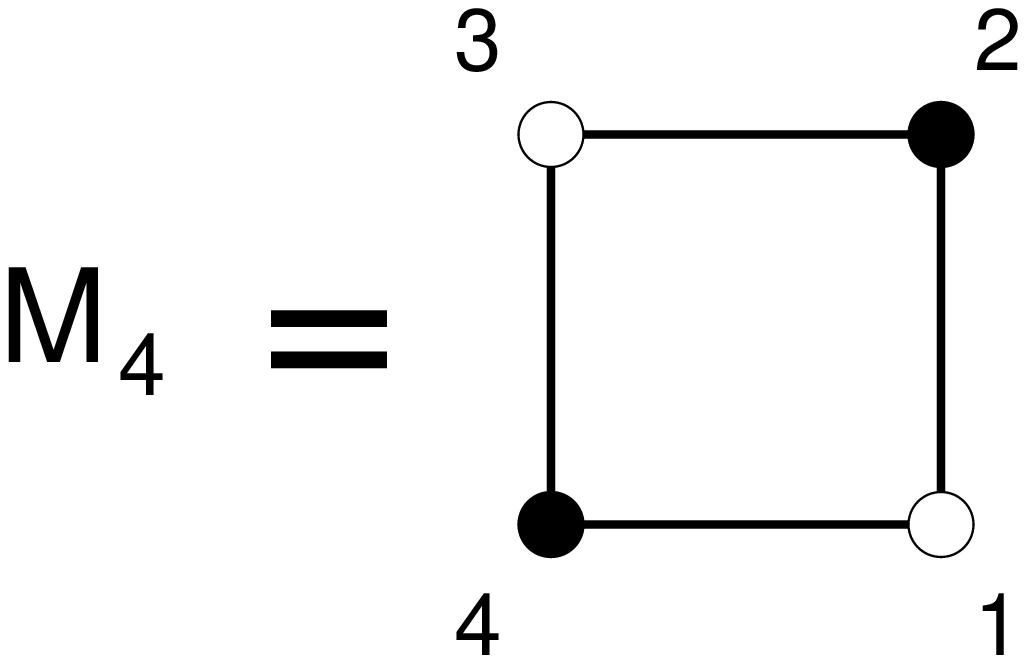} \nonumber
\ee

In discussing gravity, it is useful to introduce some further notation.
As we have seen, gravity amplitudes involve $|x| = x$ sgn $x$, so it
makes sense to introduce a separate notation for the $``x"$ factor,
as distinct from the ``sgn" factors. We denote these by red lines:

\be
\includegraphics[scale=0.5]{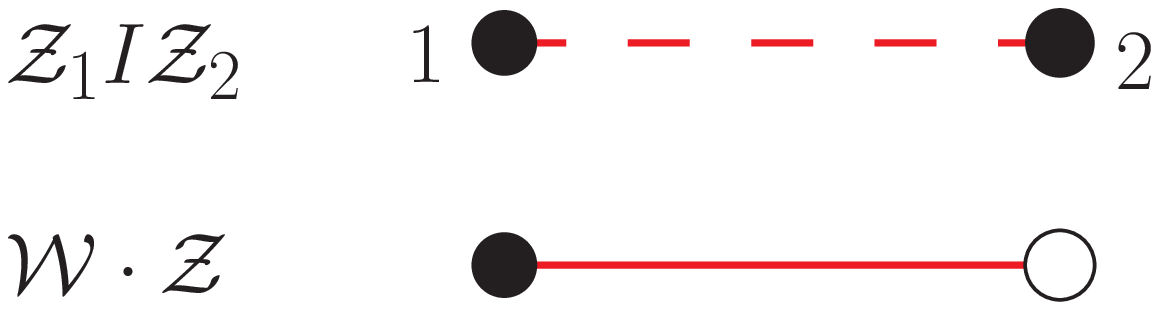} \nonumber
\ee

With this notation, the Hodges diagram for the three point functions $M_3^{+},M_3^{-}$ in ${\cal N} = 8$ SUGRA are shown below:

\be
\includegraphics[scale=0.6]{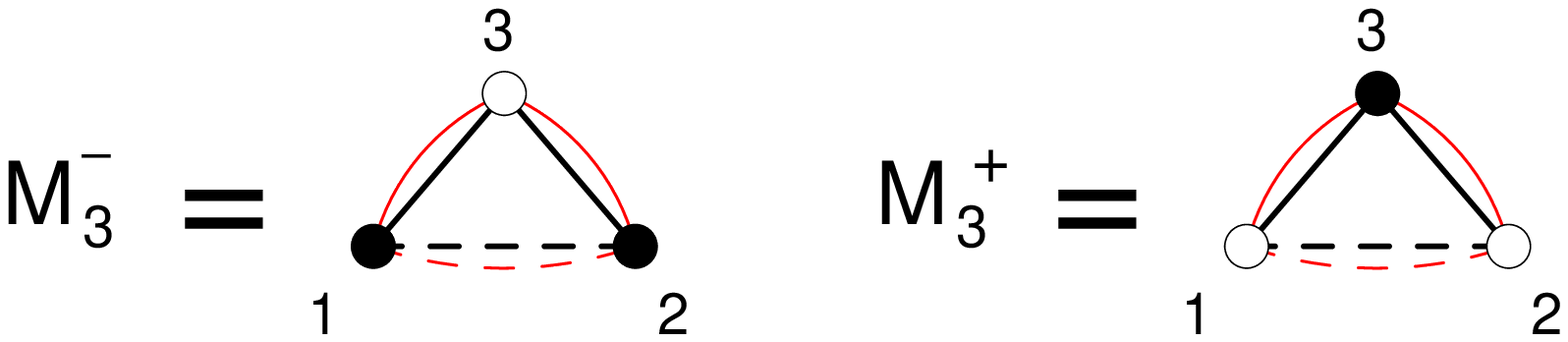} \nonumber
\ee


\subsection{The BCFW Bridge}

The BCFW recursion relation has a simple diagrammatic
interpretation. With maximal SUSY, there are actually two different
forms of the formula, depending on whether we use $({\cal Z},{\cal W})$ or $({\cal W},{\cal Z})$
variables to label the internal particle that is to be integrated
over. For ${\cal N} = 4$ SYM, we show both forms of the BCFW
``bridge" in the figure:

\be
\includegraphics[scale=0.75]{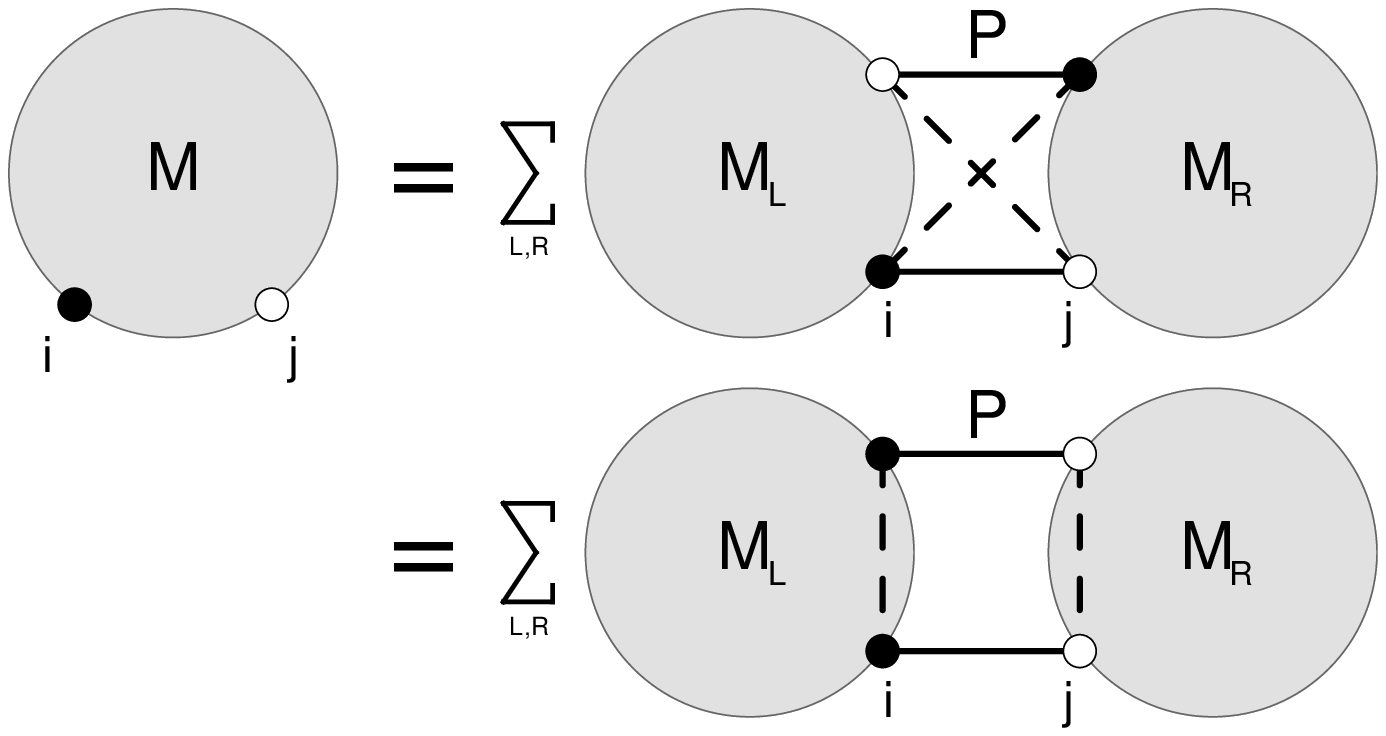} \nonumber
\ee

For ${\cal N} = 8$ SUGRA, the two forms of the BCFW bridge are
\be \label{3/4}
\includegraphics[scale=0.75]{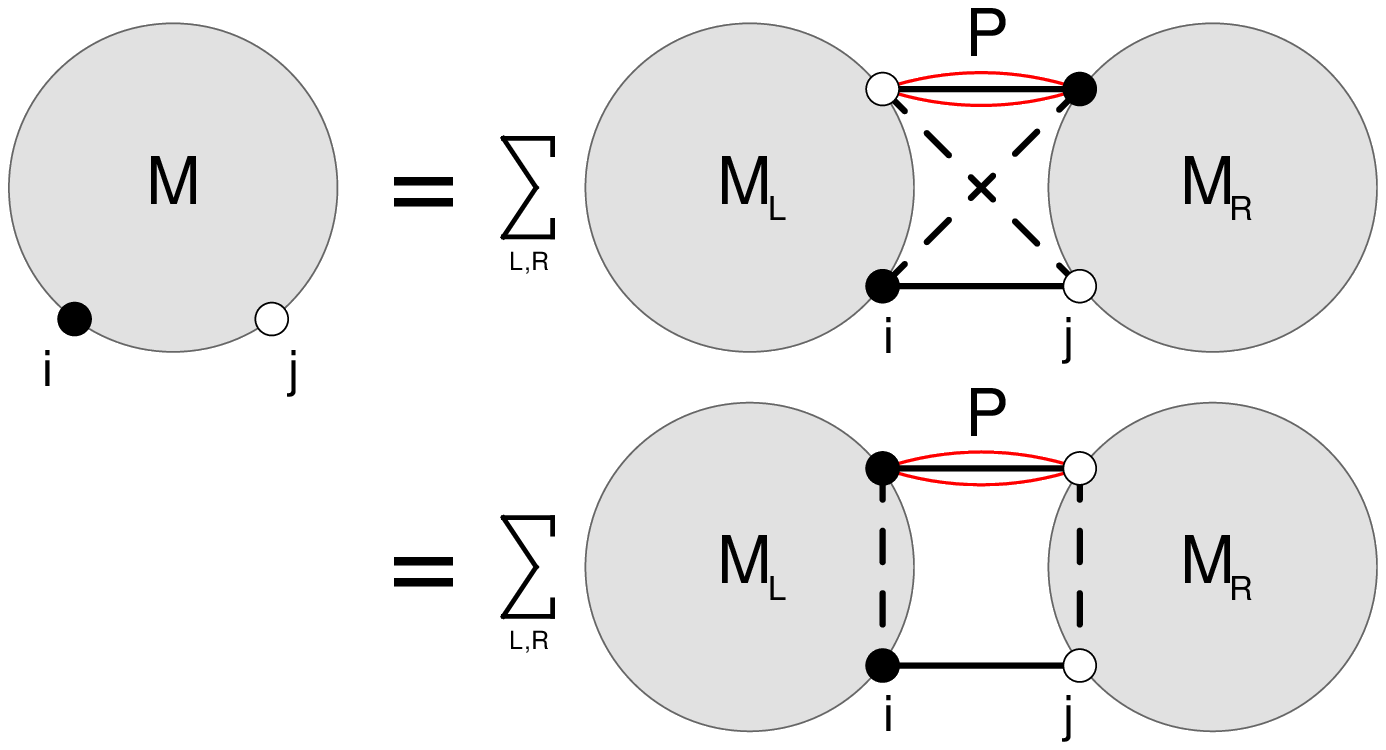} \nonumber
\ee
 Note a crucial fact. As we have drawn it, the three-point
amplitude looks like a disk, not a ``tree". Because of the
sgn${\cal W}_i \cdot {\cal Z}_j$ factor, the BCFW bridge builds
higher-point amplitudes to have the topology of a disk and not
trees! We will see this explicitly in the examples below. Having
tree-diagrams be represented by diagrams that look like disks is
very suggestive of an open string theory in
twistor space underlying ${\cal N} = 4$ SYM, which is perhaps not
surprising given the success of Witten's twistor string theory at
tree level. But it is interesting that it is the structure of the
BCFW diagrams (and not the CSW diagrams directly associated with
Witten's twistor string) that seems to be calling for an
open-string intepretation.

\subsection{Computing SYM Amplitudes With Hodges Diagrams}

Let us now use this notation to illustrate the computation of
higher-order amplitudes using the BCFW rules and Hodges diagrams
in ${\cal N} = 4$ SYM. Let us first determine what the full 3
point amplitude $M_3 = M_3^{+} + M_3^{-}$ looks like; as we have
seen $M_3^{+}$ is simple in the ${\cal W} {\cal W} {\cal Z}$ basis
while $M_3^{-}$ is simple in the ${\cal Z} {\cal Z} {\cal W}$
basis. However, we know that in the, say, ${\cal W}_1,{\cal
W}_2,{\cal W}_3$ basis, the 3 point amplitude must be fully
cyclically symmetric. This leads to the first of a series of
identities that will make it easy to manipulate twistor diagrams,
shown below, that we call ``the triangle identity":

\begin{equation} %
\includegraphics[scale=0.77]{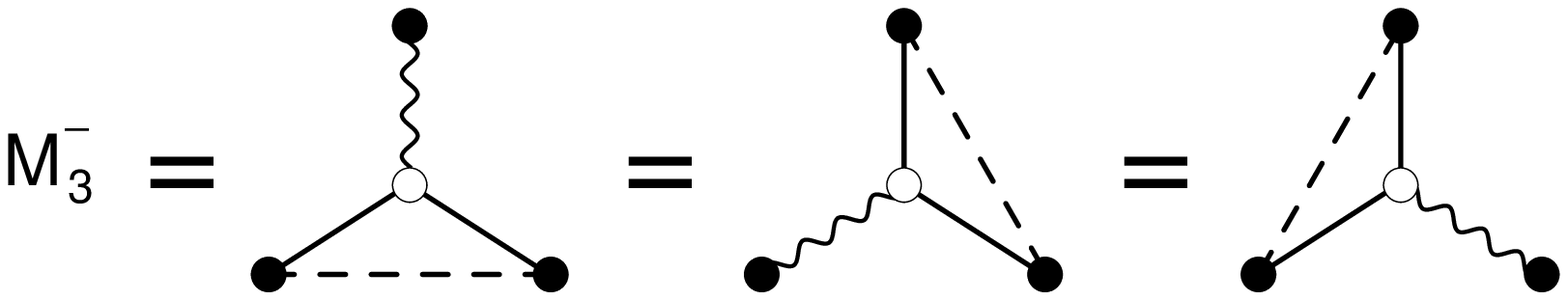} \nonumber
\end{equation}
This is a good place to mention the ``square identity", which reflects both parity invariance and the cyclic invariance of the 4-point amplitude:

\be
\includegraphics[scale=0.6]{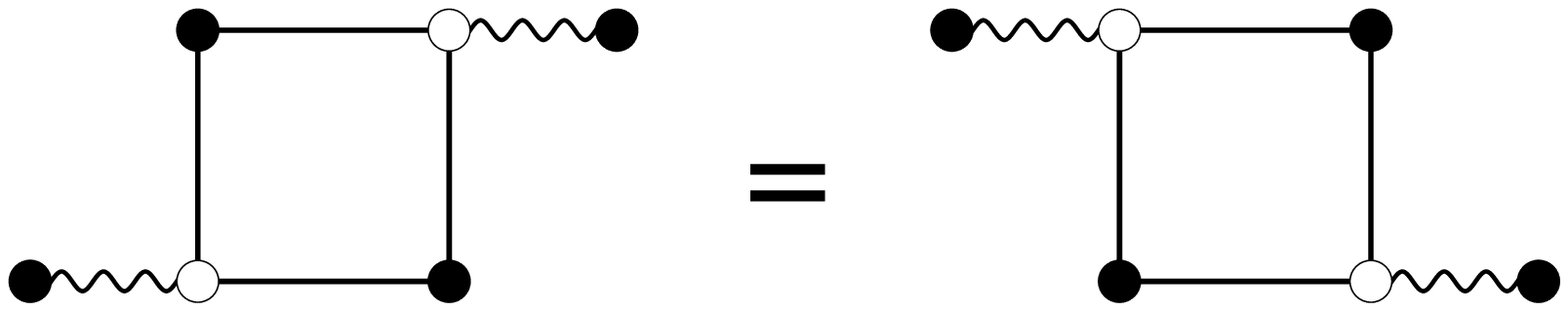} \nonumber
\ee
In both of these pictures, the white dots are to be integrated over. Obviously we can write these identities in a number of different bases as well, by twistor transforming some of the external dots; for instance another form of the square identity is

\be
\includegraphics[scale=0.6]{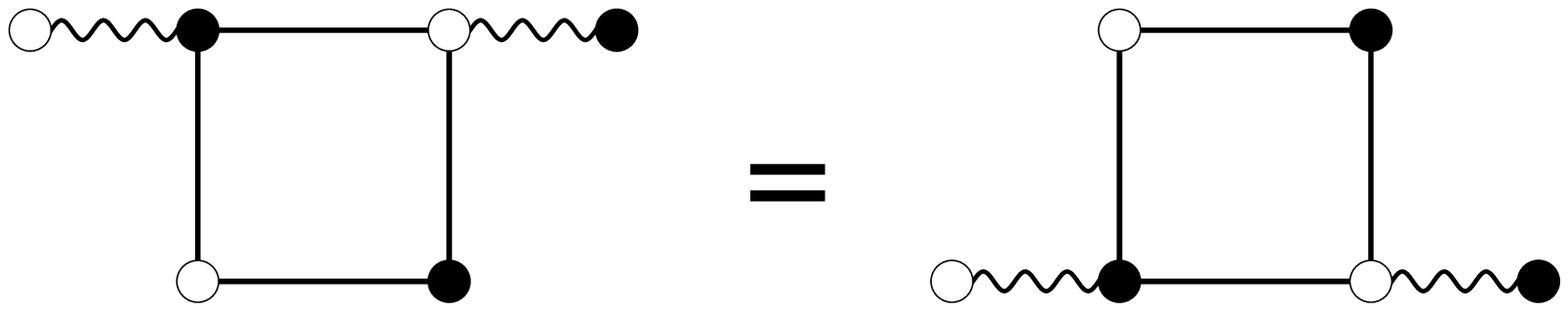} \nonumber
\ee
where the internal dots connected to the squiggly lines are integrated over.

Using the triangle identity, $M_3 = M_3^{+} + M_3^{-}$ is given in the ${\cal Z},{\cal Z},{\cal W}$ basis as

\be
\includegraphics[scale=0.6]{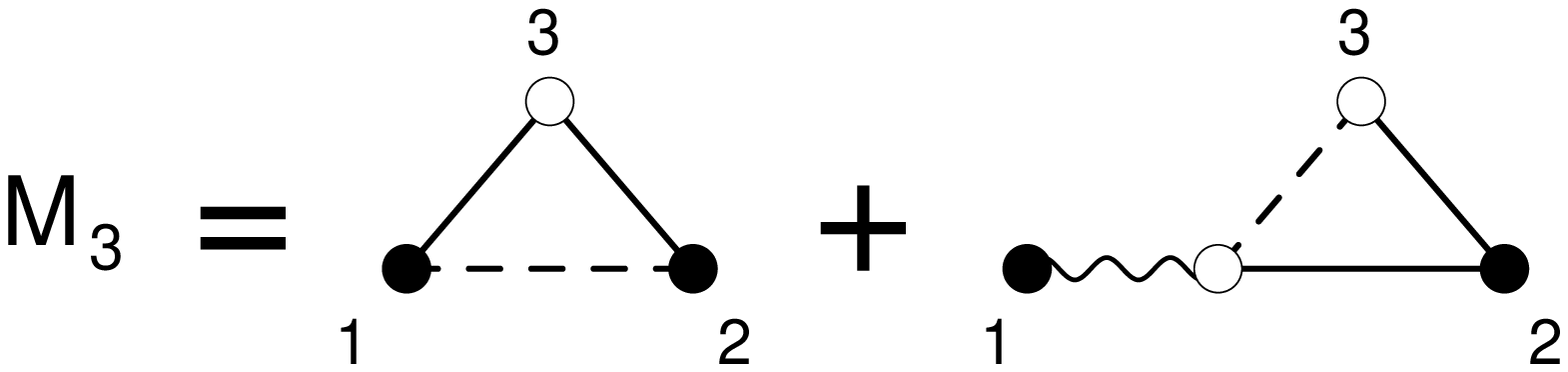} \nonumber
\ee

Let's now look at the 4-point amplitude. Now, without SUSY, there
is only one internal configuration of helicities for the internal
line that contributes to the 4 point amplitude; however with
maximal SUSY, since the 3-point amplitude is the sum of two terms,
we would appear to have $2 \times 2 = 4$ terms to keep track of.
Fortunately there is a very simple ``vanishing" identity:
\be
\includegraphics[scale=0.6]{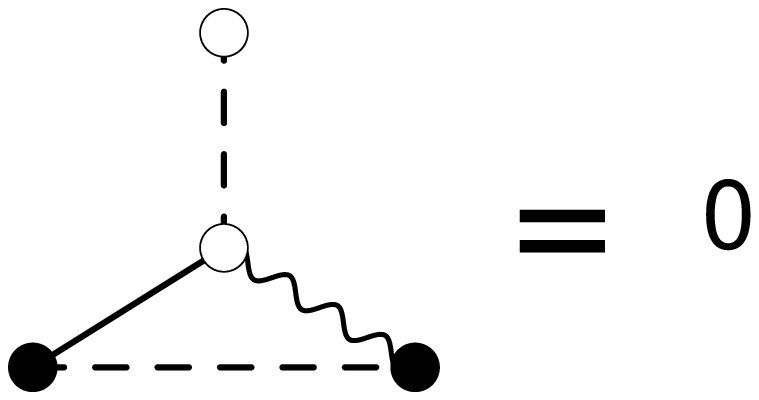} \nonumber
\ee Here the middle white dot is to be integrated over. This
identity will, in a fully supersymmetric way, enforce that only
one term contributes in the BCFW computation of the 4pt function.

We can now get on with the business of carrying out the projective
integrals in the BCFW formula. There are two identities that allow
us to de-projectivize the integrals in an extremely useful way. The
``scrunch" identity is simply a projective version of doing a
fourier-transform followed by an inverse fourier transform. The
``butterfly" identity handles a structure that will appear
ubiquitously in the BCFW bridge. These identities are
straightforward to derive directly, though we will shortly give them
a transparent motivation and proof. \be
\includegraphics[scale=0.7]{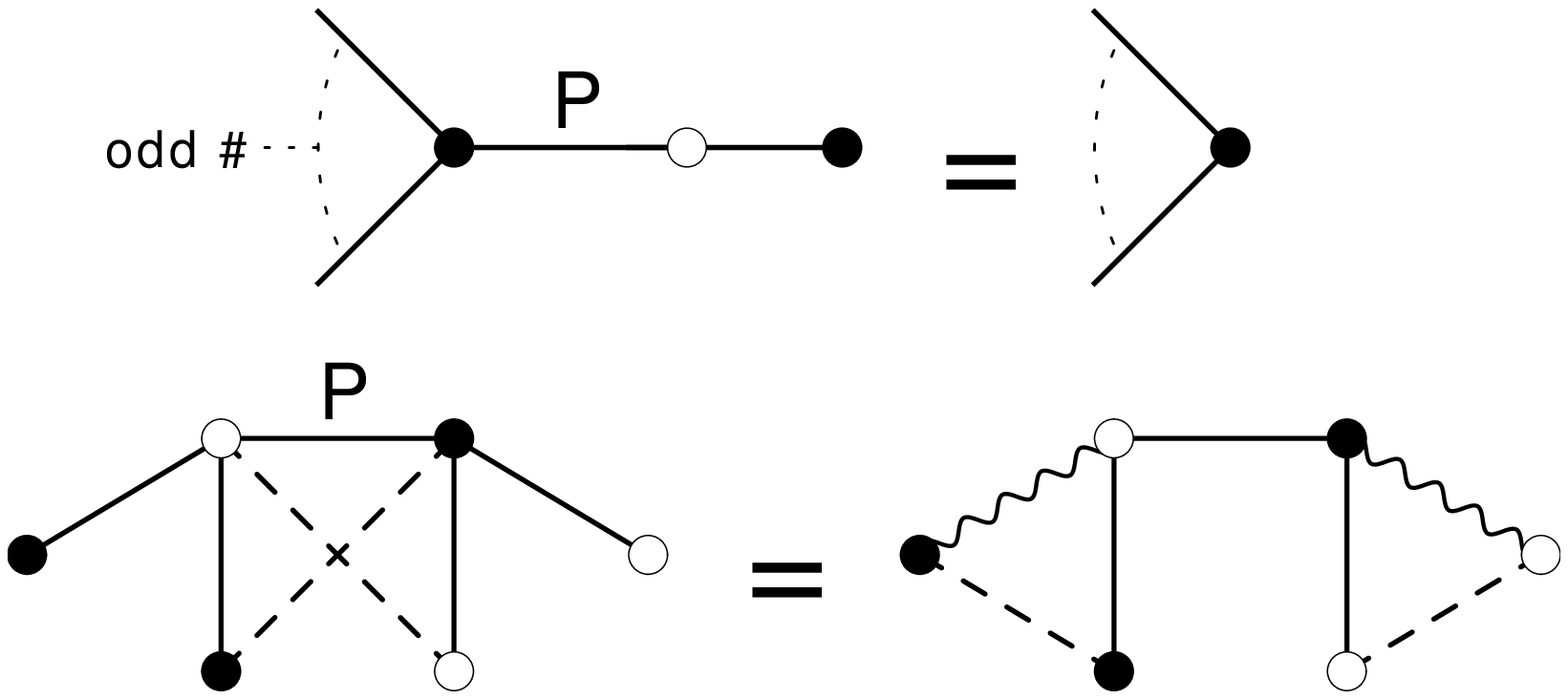} \nonumber
\ee
In these figures the ``P" denotes a projective integral, and the dots attached to the line marked with the ``P" are being integrated over. In the scrunch identity, an odd number of connections are needed for the projective integral to be well-defined.

We can finally compute the 4-pt amplitude.  We illustrate this using Hodges diagams, deforming particles 1 and 4, for both possible choices of the BCFW bridge.

\be
\includegraphics[scale=0.9]{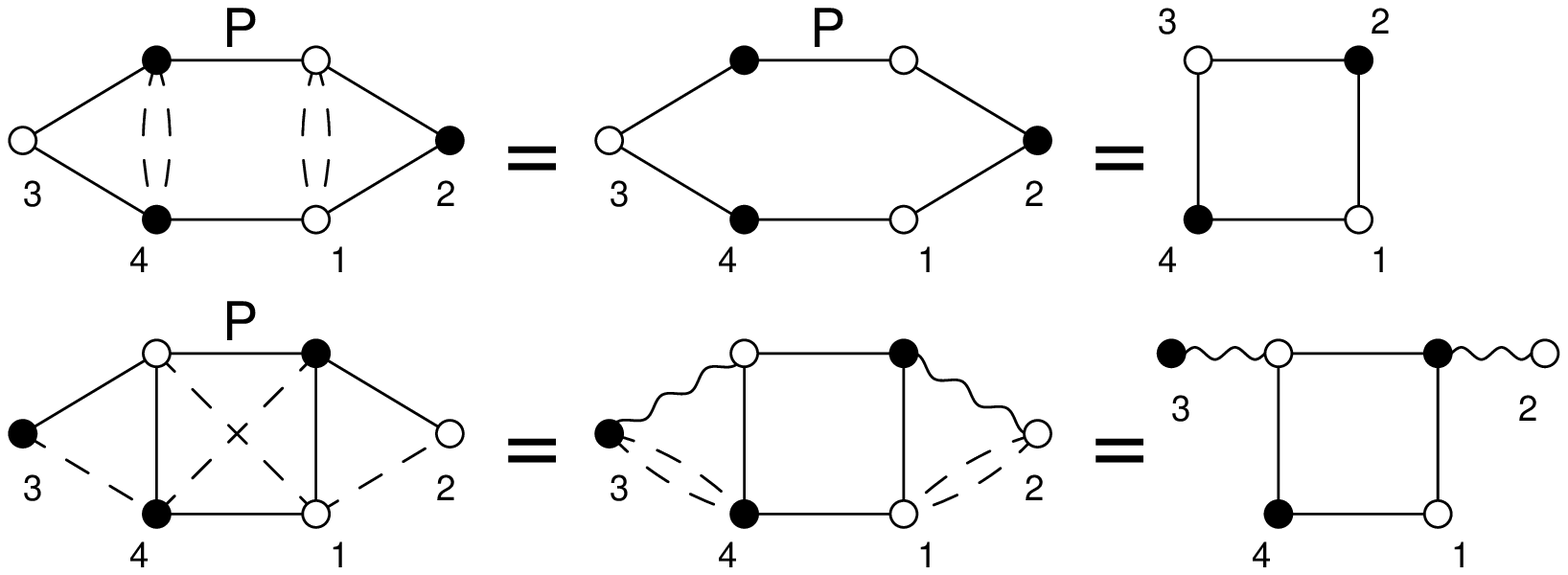} \nonumber
\ee
Note that we have deliberately not denoted the deformed particles with a hat $\hat{1},\hat{4}$, as is customary in BCFW computations. This is to emphasize that in twistor space the variables are not deformed!
In both cases, due to the vanishing identity, only one term from $M_{3L,R}$ contributes. In the first case,
in the first step we use the identity sgn$^2(x) = 1$, which can be used everywhere in these computations since sgn$^2(x)$ is integrated against functions without $\delta(x)$ type singularities. We then use the scrunch identity to bring the Hodges diagram to the form of the correct answer. For the second BCFW bridge, the butterfly identity is used, again in conjunction with sgn$^2= 1$. Indeed, given that we independently knew the 4-pt amplitude from direct fourier transformation, the scrunch and butterfly identities can be motivated and proved by matching the known amplitude to its BCFW construction.

We can see more explicitly that the other three terms vanish due to the vanishing identity:
\be
\includegraphics[scale=0.77]{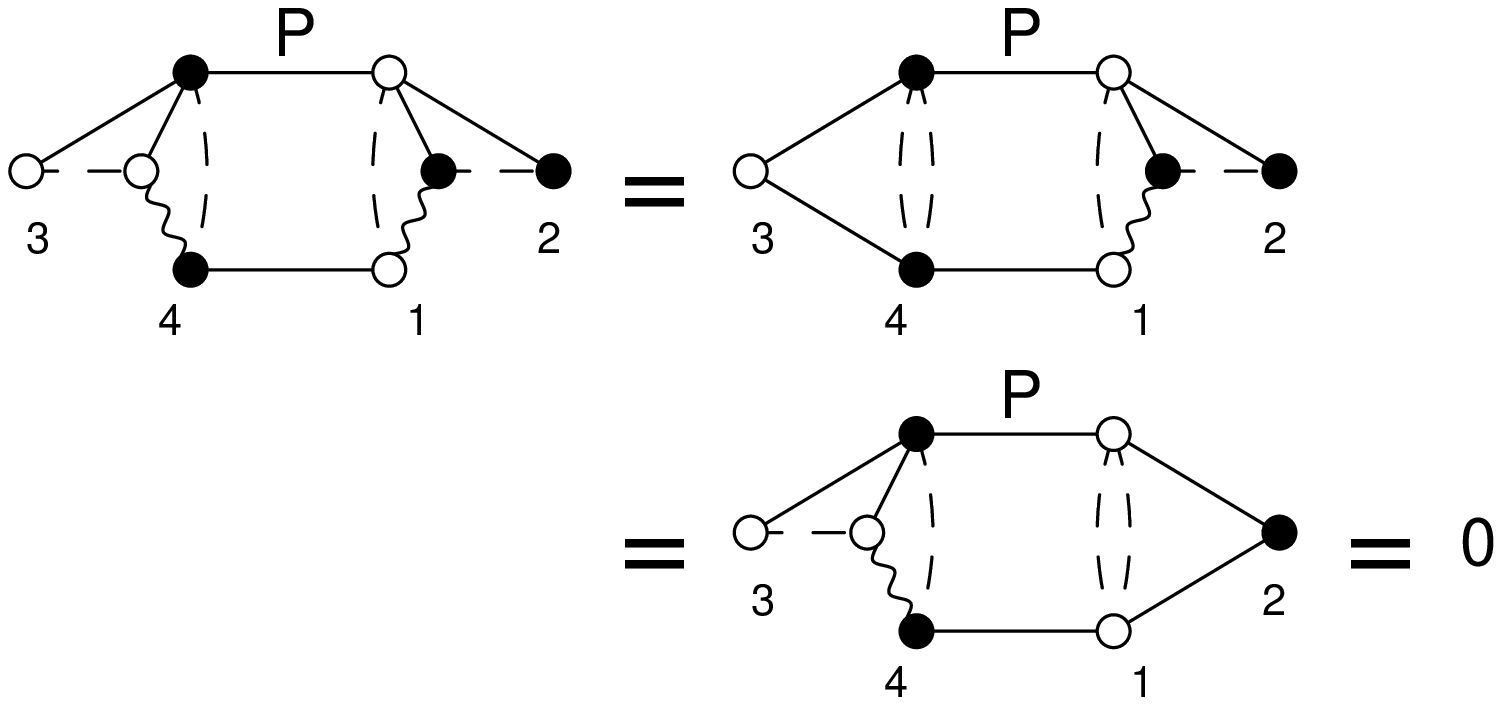} \nonumber
\ee

Using these identities, we can easily compute higher-point YM amplitudes without ever touching an explicit integral. For instance the Hodges diagrams for the 5-pt $\overline {\rm MHV}$ amplitude and the 6-pt NMHV amplitude are shown below:


\be\label{5/6}
\includegraphics[scale=0.6]{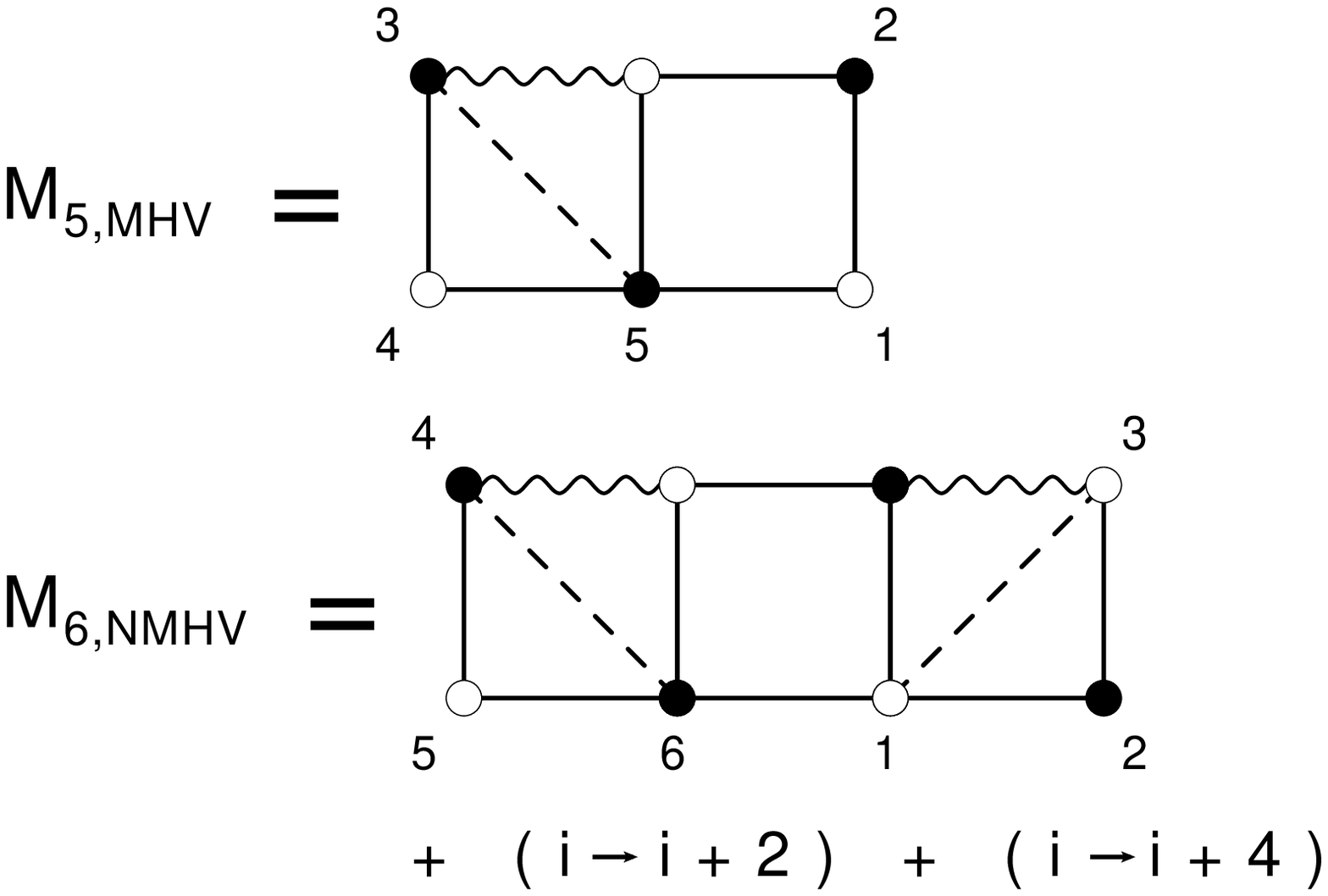} \nonumber
\ee We show the 5pt ${\rm MHV}$ computation below, choosing 1 and
5 to be the reference particles, and only showing the terms from
the 3pt amplitude that survive the vanishing identity:
\be
\includegraphics[scale=0.675]{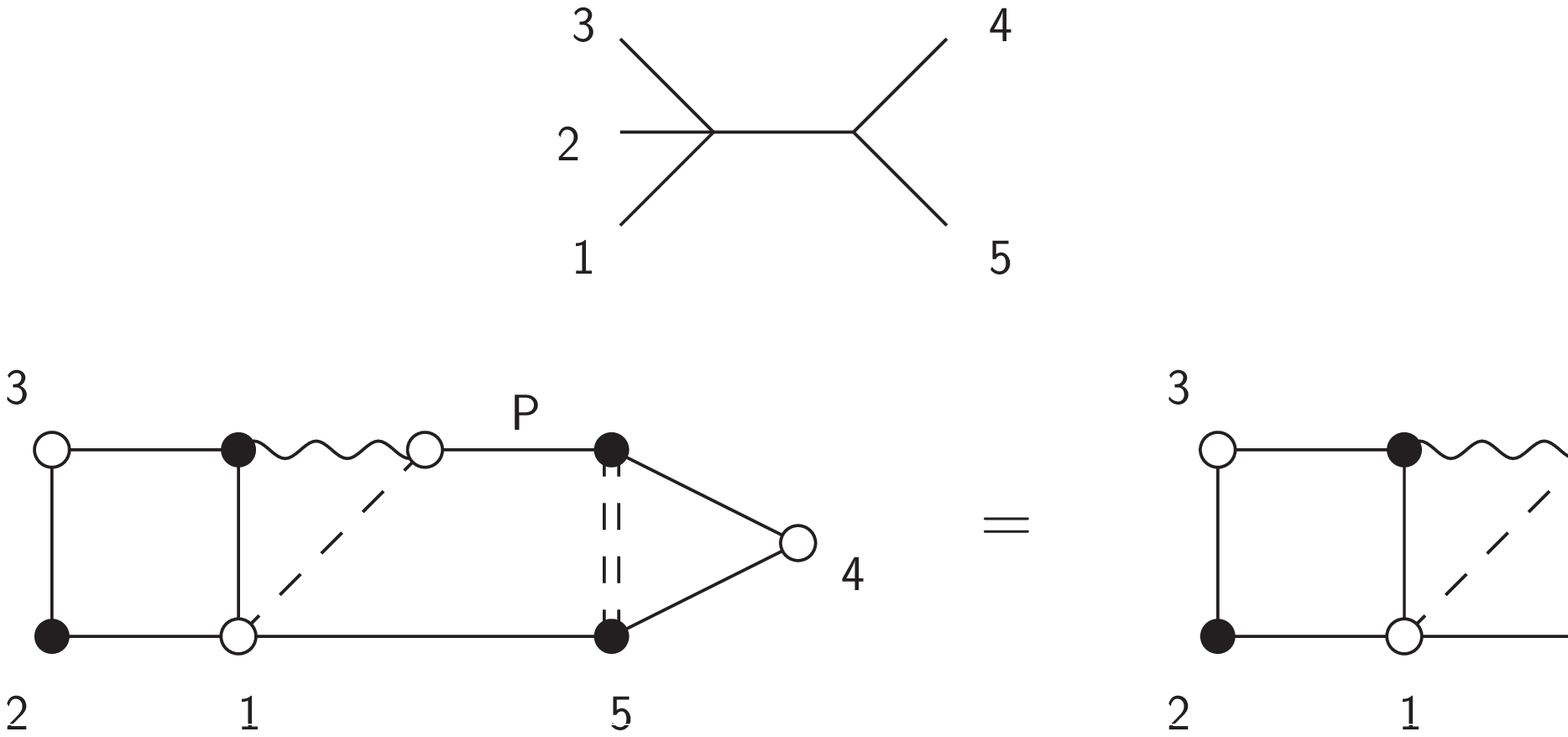} \nonumber
\ee
Here we use again that sgn$^2 = 1$, and the scrunch identity.

Next, let us compute the 6pt NMHV amplitude. Choosing 1 and 6 as
the reference particles, we first consider the term involving the
product of two 4-point functions:
\be
\includegraphics[scale=0.675]{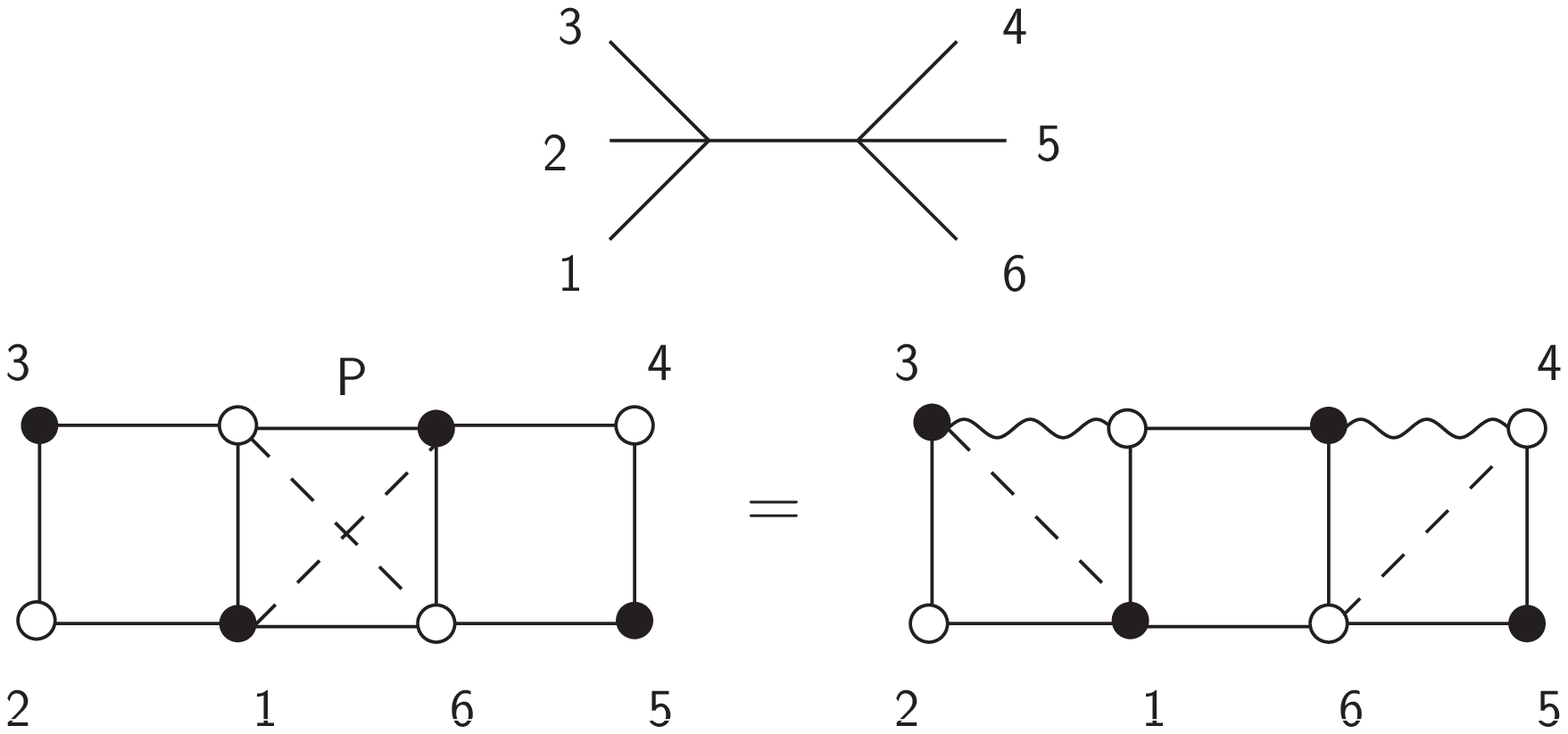} \nonumber
\ee Here we used the butterfly identity to de-projectivize the
integral in the BCFW bridge. Now let us look at the contribution
from the term involving the product $M_{3L} M_{5R}$:

\be
\includegraphics[scale=0.675]{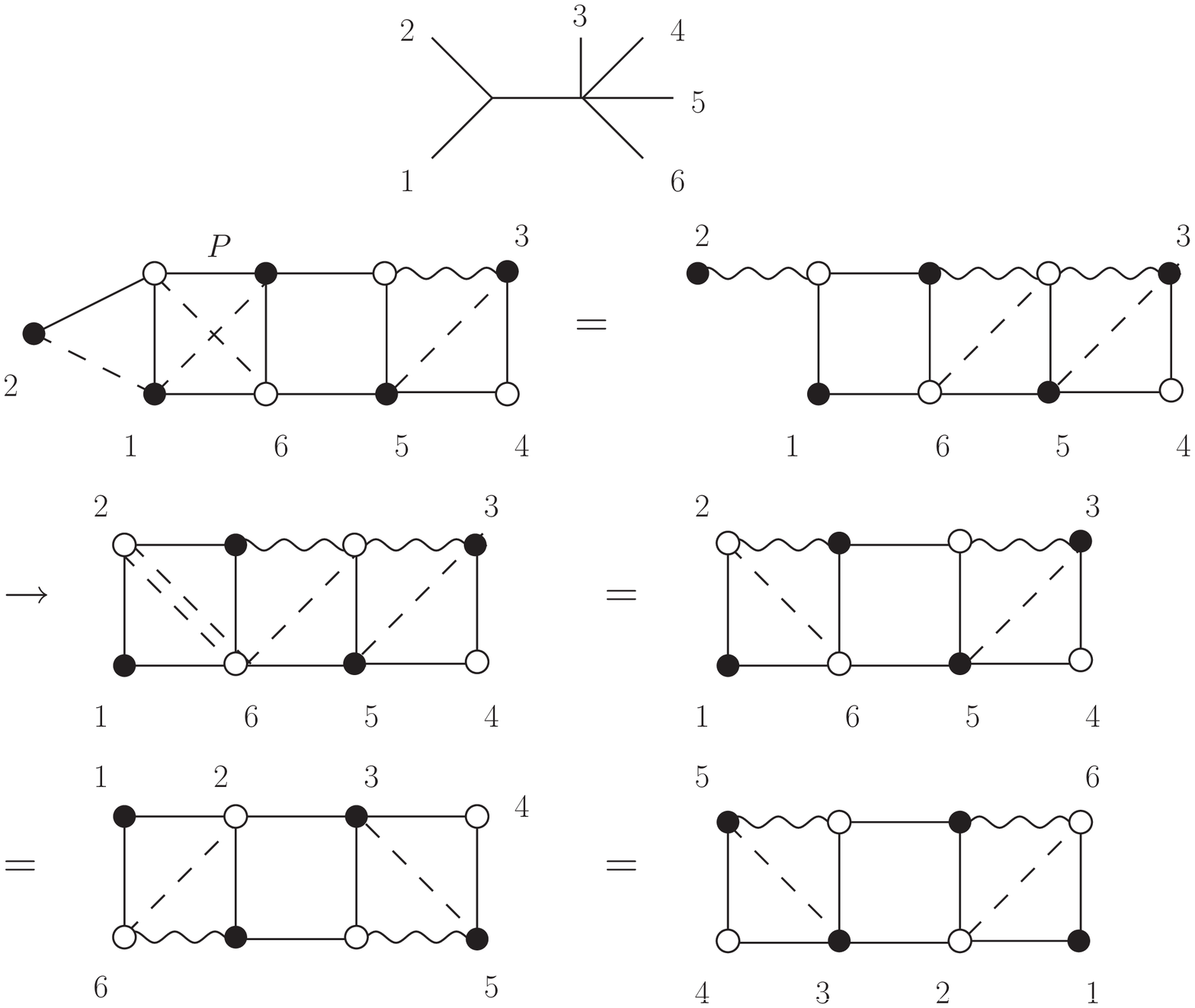} \nonumber
\ee

This time both pieces of $M_{3L}$ do contribute, but one term only contributes to the 6 pt MHV amplitude; we will only look at the term contributing to the NMHV amplitude.
In the first line we use the butterfly identity and sgn$^2 = 1$. The hanging wiggly line connecting to the black 2 dot tells us it is more natural to use a white dot to represent 2. The diagram still looks rather asymmetrical, but we can make it look nicer but introducing $1 =$sgn$^2{\cal W}_2 I {\cal W}_6$. We are thus led to the second line. We can now use the triangle identity on the triangle made of 2,6 and the internal black dot. Again using sgn$^2= 1$, this leads us to the second diagram on the second line. We could in principle stop here, but it amusing and useful to see other forms of this object that can be obtained by applying the square identity on the middle square; this is the first equality on the third line. The last figure is simply rotating the figure to make it look like the other term we've seen, with the wiggly lines on the top!

Note that with these diagrammatic manipulations, we have already
discovered something very interesting: the contribution from
$M_{3L} M_{5R}$ is of precisely the same form as $M_{4L} M_{4R}$,
with all the indices shifted by 2, $i \to i+2$! By symmetry, the
contribution from $M_{5L} M_{3R}$ must be the same with $i \to i
-2$ or what is the same $i \to i + 4$. This is not at all obvious
from the BCFW formula itself! If we were to set the Grassman
parameters to zero to obtain the 6pt alternating helicity
amplitude, we would expect it to have this cyclic symmetry, but
there is absolutely no reason to expect that the three BCFW terms
would be related to each other in this way;  indeed this fact
comes as a surprise in the explicit momentum space
calculation~\cite{BCF}. But it is made obvious with the Hodges
diagrams in conjunction with the square identity, which as we
mentioned enforces the parity invariance of the 4pt amplitude.

Note that in the 5 and 6
pt examples we have discussed, we used one form of the BCFW bridge; of course we could
have also used the other form; indeed the 6 pt computation can be done even more quickly in this way,
as the interested reader can easily verify.

There are a similar set of manipulations for ${\cal N} = 8$ SUGRA; we will only give the Hodges diagram for MHV 5 point and NMHV 6 point amplitudes as an illustration:
\begin{equation} %
\includegraphics[scale=0.6]{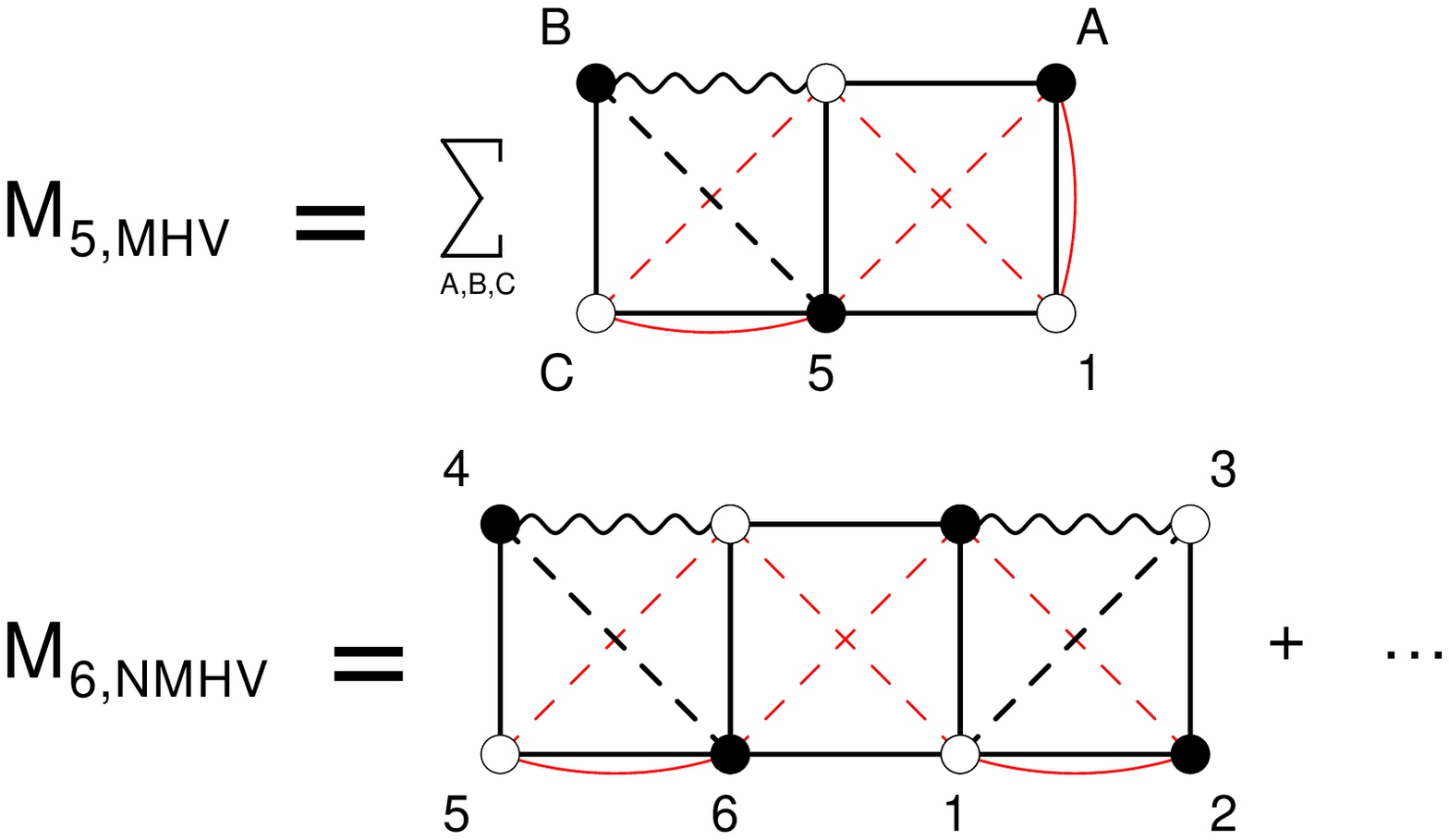} \nonumber
\end{equation}
The $+ \cdots$ indicates the sum over all permutations; note that here different terms in the BCFW sum are naturally given with different ${\cal W},{\cal Z}$ assignments.

\subsection{$M^{+-+-+-}$ from its Hodges diagram}

As an illustration of the power of these techniques, let us compute the 6 particle NMHV amplitudes back in momentum space. As we just mentioned, by simply looking at the Hodges diagrams we can see the non-trivial fact that the NMHV amplitude is the sum of three terms that are related to each other by shifting the particle labels by two units: $M^{{\rm NMHV}}_{6}= (1 + g^2 + g^4) U^{{\rm NMHV}}_6$, where $g$ is the operation that shifts the particle labels by 1, i.e.,  $g: i \to i+1$, and we can take $U^{{\rm NMHV}}$ to correspond to, say, the $M_{4L} M_{4R}$ term. Let us compute $U_6^{{\rm NMHV}}$ by looking at its Hodges diagram:
\be
\includegraphics[scale=0.6]{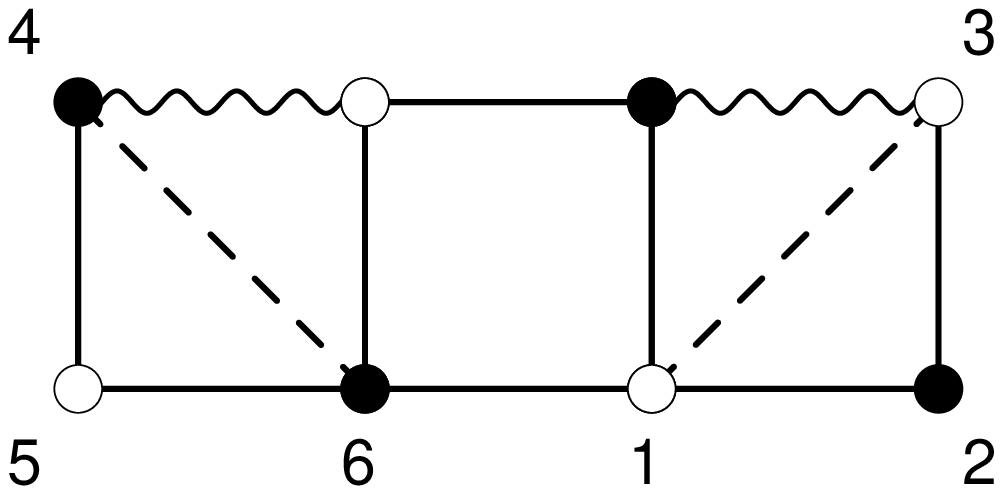} \nonumber
\ee

We may at first be put off by the internal points that need to be
integrated over. However, note that these only serve to represent
the $(6431)$ amplitude in the ${\cal Z}_6 {\cal Z}_4 {\cal W}_3
{\cal W}_1$ representation. Indeed by inspection we see that
\begin{eqnarray}
M_{6}^{{\rm NMHV}} = M_3^{-}({\cal Z}_4, {\cal W}_5, {\cal Z}_6) \times  M_4({\cal Z}_4, {\cal Z}_6, {\cal W}_1, {\cal W}_3) \times M_3^{+}({\cal W}_1, {\cal Z}_2, {\cal W}_3) \nonumber \\ + \left[i \to i + 2\right] + \left[i \to i + 4\right]
\end{eqnarray}
This is a truly remarkable formula, showing that in twistor space, the six point amplitude is the sum of three terms which are each simply {\it products} of lower-point amplitudes! This fact is not at all manifest from the structure of the BCFW recursion relations, and in fact suggests an entirely new picture for determining scattering amplitudes, which we will develop at greater length in \cite{HolS}. However, for our present purposes, this expression allows us to directly determine the link representation of the six point function!  Note that we have the links $c_{54},c_{56}$ from $M_3^{-}$, $c_{12},c_{32}$ from $M_3^+$, and $c_{14},c_{16},c_{34},c_{36}$ from $M_4$, for a total of 8 = $2  \times 6$ - 4 links; we are missing $c_{52}$. Since we have the link representation for the 3 and 4 point functions (the link representation for $M_4$ in $({\cal W}, {\cal W}, {\cal Z}, {\cal Z})$ basis being given in equation(\ref{wwzz})),  we can immediately write the formula for the 6 point function; we can do this for the full super-amplitude, but let us simply set all the $\eta, \tilde \eta \to 0$ to get a link representation for the alternating helicity $M^{+-+-+-}$ amplitude,
\begin{equation}
M^{+-+-+-} = (1 + g^2 + g^4) U^{+-+-+-}
\end{equation}
with
\begin{equation}
 U^{+-+-+-} = {\rm sgn}(\langle 46 \rangle [1 3])  \int dc_{(iJ) = ({\rm odd, even}) \neq (52)}  \delta^2(\lambda_i - c_{iJ} \lambda_J) \delta^2(\tilde \lambda_J + c_{iJ} \tilde\lambda_i) \hat{U} (c_{iJ})
\end{equation}
where
\begin{equation}
\hat{U}(c_{iJ}) = \frac{1}{c_{54} c_{56}} \times \frac{1}{c_{14} c_{36} (c_{14} c_{36} - c_{16} c_{34})} \times \frac{1}{c_{12} c_{32}}
\end{equation}
Once again there is actually no integral to be done here, since
the $\delta^2$ fully determine all 8 $c_{iJ}$'s! As for the
4-particle amplitude, this link representation is in fact the most
invariant way of writing the 6 particle amplitude in momentum
space; different ways of solving for the $c_{iJ}$ and explicitly
factoring out the momentum-conserving $\delta$ function will give
different forms of the delta-function stripped amplitude. One
simple choice is to note that since there is no $(52)$ link, we
can use the $\lambda_5$ and $\tilde \lambda_2$ equations to solve
for $c_{56},c_{54},c_{12},c_{32}$,
 \begin{equation}
 c_{56} = \frac{\langle 5 4 \rangle}{\langle 6 4 \rangle}, \, c_{54} = \frac{\langle 5 6 \rangle}{\langle 4 6 \rangle}, \, c_{12} = \frac{[23]}{[31]}, c_{32} = \frac{[21]}{[13]}
 \end{equation}
and then use, say, the $\tilde \lambda_4, \tilde \lambda_6$ equations to solve for the rest of the links,
\begin{equation}
c_{14} = \frac{\langle 6 | (p_5 + p_4) |3]}{\langle 4 6 \rangle [1 3]}, c_{34} =  \frac{\langle 6 | (p_5 + p_4) |1]}{\langle 4 6 \rangle [3 1]}; c_{16} = \frac{\langle 4 | (p_5 + p_6) |3]}{\langle 4 6 \rangle [3 1]}, c_{36} = \frac{\langle 4 | (p_5 + p_6) |1]}{\langle 6 4 \rangle [1 3]}
\end{equation}
A few applications of the Schouten identity also identifies $c_{16} c_{34} - c_{14} c_{36}$ as a familiar object:
\begin{equation}
c_{16} c_{34} - c_{14} c_{36} = \frac{(p_4 + p_5 + p_6)^2}{\langle 4 6 \rangle [1 3]}
\end{equation}
Finally, it is very easy to see that the Jacobians in coverting
the $\delta^2$ integrals into single $\delta$'s fixing the
$c_{iJ}$, together with the one coming from converting the
remaining two $\delta^2$ involving $\lambda_1,\lambda_3$ into the
momentum conserving $\delta$ function, combine with the
sgn$(\langle 4 6 \rangle [1 3])$ prefactor to produce a factor of
$1/(\langle 4 6 \rangle [1 3])$. We are then left with
\begin{equation}
U^{+-+-+-} = \delta^4(\sum_k p_k) {\cal U}^{+-+-+-}
\end{equation}
where
\begin{eqnarray}
{\cal U}^{+-+-+-} &=& \frac{1}{\langle 4 6 \rangle [1 3]} \frac{1}{c_{54} c_{56}} \times \frac{1}{c_{14} c_{36} (c_{14} c_{36} - c_{16} c_{34})} \times \frac{1}{c_{12} c_{32}} \nonumber \\ & = & \frac{\langle 4 6 \rangle^4 [ 1 3 ] ^4}{[12][23] \langle 4 5 \rangle \langle 5 6 \rangle} \frac{1}{\langle 6|(p_5 + p_4)|3]} \frac{1}{\langle 4|(p_5 + p_6)|1]} \frac{1}{(p_4 + p_5 + p_6)^2}
\end{eqnarray}
which matches the correct result.

\subsection{Comparison with Hodges' Work}

We close this section by briefly discussing the connection between our work and that of Hodges. The idea of relating twistor diagrams to field theory scattering amplitudes goes back to Penrose's work in the 1970's~\cite{Penrose}. However, it was Hodges~\cite{Hodges} who, very shortly after the introduction of the BCFW recursion relations, realized their connection to twistor diagrams in an ambidextrous formalism with both twistor and dual twistor variables! With remarkable intuition, he understood the structure of the BCFW bridge, and also understood the importance of many of the identities analogous to the ones we discussed above in his formalism.

The twistor diagrams are defined as contour integrals over complex
twistor space, but the catch is that the rule for specifying the
correct contour of integration is not known. On the other hand, we
have concretely defined real integrals in $(2,2)$ signature; this
should help specify the correct contour in Hodges' formalism. The
tell-tale sign of the extra information we are adding is that our
``Hodges diagrams" are decorated with more structures than the
ones Hodges draws--the dashed lines corresponding to sgn factors
involving infinity twistors, and the wiggly lines specifying full
twistor transforms, as seen below for the 6pt NMHV amplitude: \be
\includegraphics[scale=0.8]{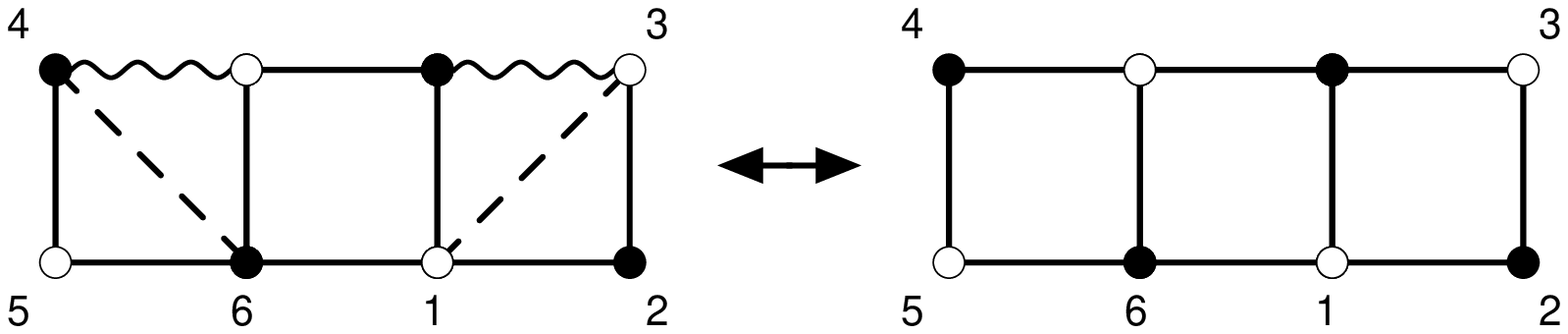} \nonumber
\ee
Note that the sgn factors in particular really can be thought of as specifying a contour of integration: the sgn's appear in jacobians because real integrals run from e.g. $- \infty$ to $\infty$ rather than the other way around.


Despite not knowing the precise contour of integration, Hodges is able to use his diagrams to impressively compute the amplitudes in Yang-Mills theory up to 8 points. How can he do this? The answer is that he also knows a BCFW-type recursion relation, not for computing amplitudes, but for computing a higher-point momentum conserving $\delta$ function in twistor space! He then arranges to act on this delta function by differential operators, to convert the $\delta$ function Hodges diagrams to the amplitude Hodges diagrams; indeed he used this idea to guess the form of the BCFW bridge. Since the action of the differential operators don't depend on the contour of integration, Hodges doesn't need to know the contour in order to be able to obtain the amplitude by acting with differential operators on the $\delta$
function.  Our use of the link representation accomplishes very much the same thing in a simpler and systematic way, reducing the computation of amplitudes to pictorial manipulations with Hodges diagrams and solving linear equations to go back to momentum space.

It would be extremely interesting to make the connection between our picture in $(2,2)$ signature with the picture using complex integration in twistor space. Not only should this allow us to make more direct contact with physics in $(3,1)$ signature, but it might also allow us to exploit the full power of Cauchy's theorem in higher dimensions in understanding the remarkable structure of twistor-space amplitudes.

\section{Tree-Level Holography for SYM and SUGRA}

The recursive form of the BCFW relation has simple analogs in elementary mathematics. For instance, the catalan numbers--which are the most ubiquitous objects in combinatorics after the binomial coefficients--are defined recursively by $C_1 = 1$ and
\begin{equation}
C_{N+1} = \sum_{i=1}^N C_{N - i} C_i
\end{equation}
which has precisely the same structure as the BCFW relations. Indeed the total number of terms in the BCFW expansion of super-amplitudes in ${\cal N} = 4$ SYM are just these catalan numbers (the number of terms to the N$^k$MHV amplitudes are the Narayana numbers). As usual in combinatorics, instead of dealing with the individual $C_N$, it is natural to use a generating function
\begin{equation}
C(x) = \sum_{N = 1}^{\infty} C_N x^N
\end{equation}
Then the recursion relation takes a very simple form
\begin{equation}
\label{catalan}
C(x) -  C(x)^2 = x
\end{equation}
In this simple case this quadratic equation can be solved $C(x) = \frac{1 - \sqrt{1 - 4x}}{2}$ and the expansion in $x$ gives an explicit form for the $C_N$.

We would like to follow the analogous steps for the scattering amplitudes in maximally supersymmetric theories. The analog of $x$ will be functions $\phi({\cal W})$ on dual twistor space or $\tilde{\phi}({\cal Z}) = \int d^{4|{\cal N}} {\cal W} e^{i {\cal W} \cdot {\cal Z}} \phi({\cal W})$ on twistor space. For Yang-Mills theory these also carry a color label.

Let us introduce a functional which has the interpretation of a twistor space ``propagator" in a general background $\phi$; for Yang-Mills we define it to be
\begin{equation}
P^{ab}\left[\phi\right]({\cal W},{\cal Z}) = \sum_n \int d^{4|4}
 {\cal W}_1 \cdots d^{4|4} {\cal W}_n \phi^{c_1}({\cal W}_1) \cdots \phi^{c_n}({\cal W}_n) M^{a b c_1 \cdots c_n}({\cal W},{\cal Z},{\cal W}_1, \cdots, {\cal W}_n)
\end{equation}
while for gravity we have a similar object without the color indices
\begin{equation}
P\left[\phi\right]({\cal W},{\cal Z}) = \sum_n \int d^{4|8}
 {\cal W}_1 \cdots d^{4|8} {\cal W}_n \phi({\cal W}_1) \cdots \phi({\cal W}_n) M({\cal W},{\cal Z},{\cal W}_1, \cdots, {\cal W}_n)
\end{equation}
If we group all the amplitudes together into a generating functional of $M\left[\phi\right]$ in the obvious way then e.g. for gravity the propagator would be given by
\begin{equation}
\label{formP}
P\left[\phi\right]({\cal W},{\cal Z}) = \frac{\delta^2 M\left[\phi\right]}{\delta \phi({\cal W}) \delta \tilde \phi({\cal Z})}
\end{equation}
Given the propagator $P$, we can determine the $n$-point amplitude by taking $(n - 2)$ functional derivatives of $P$ with respect to $\phi$.

Note that while we have defined these objects as functionals of fields defined on the full $\mathbb{R}^{4|{\cal N}}$ space, since the amplitudes have well-defined projective weights under rescaling, by writing ${\cal Z} = v {\cal Z}_P, {\cal W} = u {\cal W}_P$ and integrating over $u,v$, we are left with functionals of fields depending only on $\mathbb{RP}^{3|{\cal N}}$. Indeed, the object $M\left[\phi\right]$ is the natural analog of the boundary action in AdS/CFT. A general scattering problem in asymptotically flat space is specified  by giving some classical solutions of the free theory at infinity, and these correspond to functions of appropriate weight on $\mathbb{RP}^{3|{\cal N}}$ or its dual, and so it is natural to think of the scattering amplitudes or the propagator as a functional of this object.

Before writing the functional form of the BCFW formula, let us introduce some natural notation. Consider first any function $F({\cal W},{\cal Z})$ with weight 0 under rescaling ${\cal W},{\cal Z}$; it can be thought of as defining an inner product ${\bf F}$ on twistor/dual twistor space. Now, given two functions $F({\cal W},{\cal Z}), G({\cal W},{\cal Z})$, we can define another function of weight 0, ${\bf F} \star {\bf G}$, via the BCFW measure
\begin{equation}
(F \star G)({\cal W},{\cal Z}) = \int \left[D^{3|4} {\cal W}^\prime D^{3|4} {\cal Z}^\prime \right]_{{\cal W},{\cal Z}} F({\cal W},{\cal Z}^\prime) G({\cal W}^\prime, {\cal Z})
\end{equation}
We have the obvious analog of this formula for taking the product of two objects of any weight $p$, multiplying the measure by $({\cal W}^\prime {\cal Z}^\prime)^p$.

Furthermore, there is a natural association of a function $\Phi({\cal W},{\cal Z})$ with a function $\phi(W)$, using the three-point amplitude:
\begin{equation}
\Phi({\cal W},{\cal Z}) = \int d^{4|{\cal N}} {\cal W}^\prime M_3({\cal W},{\cal Z},{\cal W}^\prime) \phi({\cal W}^\prime)
\end{equation}
This is canonical in the sense that the three-point amplitude is essentially completely determined by specifying its projective weights.

With this notation in hand, the generating functions for maximally
supersymmetric Yang-Mills and Gravity are determined by
\begin{equation}
\label{quadYM}
{\bf P}^{ab}\left[\phi\right] - {\bf P}^{ac}\left[\phi\right] \star {\bf P}^{cb}\left[\phi\right] = g f^{a b}_{\, c} {\bf \Phi}^c
\end{equation}
and
\begin{equation}
\label{quadGRA}
{\bf P}\left[\phi\right] - {\bf P}\left[\phi\right] \star {\bf P}\left[\phi\right] = \frac{1}{M_{{\rm Pl}}} {\bf \Phi}
\end{equation}
which are the functional analogs of the quadratic equation
(\ref{catalan}) determining the catalan number generating function
$C(x)$. We have restored the coupling constant dependence on the
right-hand side to show how the three-point amplitude acts as the
``source term" in these equations, forcing $P$ to be
non-vanishing\footnote{It is easy to show that $1\star 1 = 0$, and
so $P$ vanishes in the absence of a source.}, and also completely
determining it. These strikingly simple equations can be thought of
giving a completely holographic definition of ${\cal N} = 4$ SYM and
${\cal N} = 8$ SUGRA at tree level.

While this holographic formula makes no reference to the bulk
$(2,2)$ spacetime, it comes as close as possible to making local
spacetime physics manifest, by making obvious the recursive
determination of the scattering amplitudes, which in turn reflect
factorization in space-time. However making locality as obvious as
it can be comes at the price of not manifesting two other
important symmetries of the scattering amplitudes. One is manifest
Parity, which is broken in the BCFW formalism. The other is the
fact that we get the same amplitudes no matter which pair of
particles we choose as references! This is a highly non-trivial
fact. Indeed, note that we could write down the analog of our
quadratic equation for particles of any spin, or what is the same,
we could define amplitudes for any theory using the BCFW formulas,
starting from the (uniquely fixed) three-particle amplitude.
However, with the exception of Yang-Mills and Gravity, we will not
find that we get the same answers for different choices of BCFW
reference particles. In terms of our generating functions, this is
reflected in the fact that the solutions of the quadratic
equations (\ref{quadYM},\ref{quadGRA}) for the propagators $P$
actually take the form given in equation (\ref{formP}). Said
another way, what is special about Yang-Mills and Gravity is that
the solution of the quadratic equations for $P$ automatically
satisfy the constraint
\begin{equation}
\label{integral}
\frac{\delta P\left[\phi\right]({\cal W},{\cal Z})}{\delta \phi({\cal W}^\prime)} - \frac{\delta P\left[\phi\right]({\cal W}^\prime,{\cal Z})}{\delta \phi({\cal W})} = 0; \, \, \frac{\delta P\left[\phi\right]({\cal W},{\cal Z})}{\delta \tilde \phi({\cal Z}^\prime)} - \frac{\delta P\left[\phi\right]({\cal W},{\cal Z}^\prime)}{\delta \tilde \phi({\cal Z})} = 0
\end{equation}
Thus the equations (\ref{quadYM},\ref{quadGRA}) should be thought of as giving a holographic definition of the theory that is closest to making contact with local bulk spacetime physics, but which does not make manifest either parity or the remarkable property of equation (\ref{integral}). It is then clearly desirable to complete the transition to a holographic description that makes {\it all} these properties manifest, at the expense of losing any direct connection to spacetime locality, a topic we will explore at greater length in \cite{HolS}.

\section{Structure of the S-Matrix at Loop Level}

At tree level scattering amplitudes are rational functions of the
basic lorentz invariants constructed out of the spinors
$\{\lambda_a,\tilde\lambda_{\dot a}\}$ of the external particles.
This makes the continuation from one signature of spacetime to
another a trivial procedure for everything but the three-point
amplitude which vanishes in lorentzian signature but not in split
signature. There is however one subtlety which makes the previous
statement not fully correct. For generic values of the external
momenta, i.e., away from singularities, one can ignore any
$i\epsilon$ prescription but near poles one has to be careful and
different signatures might require different prescriptions.

In previous sections we learned that in transforming into twistor space using split signature, an integration over external momenta must be done and therefore singular points must be included. We found that in order to have a well defined action of the little group in twistor space the most natural prescription for defining the fourier transform of distributions of the form $1/x$ is the principal value prescription. This is not merely a replacement of the Feynman $i\epsilon$ prescription of propagators by the principal value as discussed in section 3.1 for the three-point amplitude.

Having to deal with integrations over momentum variables is also unavoidable at higher orders in perturbation theory regardless of twistor space. This is indeed where one finds that a continuation from one signature to another might be subtle. In fact, we will discover that the continuation from lorentzian to split signature is especially subtle due to the intricate structure of singularities present in the latter. Moreover, a full analytic continuation, analogous to a Wick rotation connecting lorentzian to Euclidean, is not available due to the presence of low codimension singularities.

In this section we set to explore these issues and try to define one-loop amplitudes in split signature. Without the notions of causality and unitarity of its lorentzian counterpart, we don't have a fundamental definition of the split signature scattering amplitudes.
We therefore take as a definition of scattering amplitudes at one loop the same set of Feynman diagrams as in the lorentzian case. The usual reduction procedures also apply in split signature, leading to a form of the amplitudes in terms of a linear combination of scalar integrals with coefficients that are rational functions.

In this first exploration we choose to concentrate on one-loop
amplitudes in ${\cal N}=4$ SYM. In particular we study in detail the
four-particle amplitude~\cite{oneloopfour}. This amplitude consists of a scalar box
integral with all massless external legs~\cite{alloneloop}. This is clearly UV finite
but it has IR divergences. We study carefully the IR divergences in
the lorentzian case, reviewing how collinear and soft singularities
appear. This allows us to pinpoint exactly what happens when the
integral is defined in split signature. We find that the same
divergences present in the lorentzian case are present but in
addition there are new divergences! Even more surprising is the fact
that all divergences, old and new, are very easy to regulate. In
fact, any $i\epsilon$ prescription regulates the integrals. This is
in sharp contrast to the lorentzian case where no $i\epsilon$
prescription completely regulates the integral. Inspired by our
tree-level discussion we take the principal value prescription as
our way to define the loop integrals in split signature.

\subsection{IR Divergences in Lorentzian and Split Signatures}

It is worth recalling how IR divergences appear in the lorentzian case. This is also a good point to formally introduce our object of study, i.e., the single trace contribution to one-loop amplitudes in ${\cal N}=4$ SYM~\cite{oneloopfour}. These one-loop amplitudes can be written as a sum over scalar box integrals times coefficients which are rational functions of the kinematical invariants. More explicitly, one has for four-particle amplitudes
\begin{equation}
\label{ampi}
M^{1\hbox{-}{\rm loop}}_4 = {\cal M}^{\rm tree}_4 st I_4(s,t)
\end{equation}
where the scalar integral $I_4$ is defined, in $D=4-2\epsilon$ dimensions, as follows
\begin{equation}
\label{four}
I_4(s,t)  =  \int d^DL \frac{\delta^4(k_1+k_2+k_3+k_4)}{L^2(L+k_1)^2(L+k_1+k_2)^2(L+k_1+k_2+k_3)^2}.
\end{equation}

This integral is clearly UV finite but it has IR divergences produced at four different isolated points in the integration region. Note that in order to get a divergence we must impose at least three inverse propagators to vanish. In lorentzian signature this is also the maximum possible number. Consider $I_4$ near the region where $L\sim 0$. Using momentum conservation to write $k_1+k_2+k_3=-k_4$ one finds that
\begin{equation}
\label{IR}
I_4(s,t)|_{\rm IR} \sim  \frac{1}{s}\int d^DL \frac{1}{L^2(L+k_1)^2(L-k_4)^2}\delta^4(k_1+k_2+k_3+k_4) \sim s^{-2-\epsilon}\frac{1}{\epsilon^2}.
\end{equation}
This result can easily be checked by using Feynman parameter methods. The IR singular behavior is the same as the one of our original integral coming from the $L\sim 0$ region. Similarly, by a change of variables, one finds analogous results from the region of integrations near $L=k_1$, $L=k_1+k_2$ and $L=-k_4$.

We must therefore regulate these integrals in the IR; dimensional regularization is the standard regularization used in actual computations of amplitudes for practical purposes~\cite{sbook} but it obscures some of the important physics which allows us to move to split signature. This is why we study the same integral (\ref{IR}) but regulated  by adding a small mass $m^2$ to the $L^2$ propagator and setting $D=4$. In this case, one can easily check, using Feynman parameters, that~\cite{sbook},
\begin{equation}
\label{IRmass}
I_4(s,t)|_{\rm IR} \sim  \frac{1}{s}\int d^4L \frac{1}{(L^2-m^2)(L+k_1)^2(L-k_4)^2}\delta^4(k_1+k_2+k_3+k_4) \sim \frac{1}{s^2}{\rm log}^2\left(\frac{m^2}{s}\right).
\end{equation}

Instead of using Feynman parameters let us perform the same computation by using a method which makes the nature of the divergences very transparent. Let us start with (\ref{IRmass}) written in light cone coordinates $(L_+,L_-,L_t)$ such that $2L\cdot k_1 = L_+$ and $2L\cdot k_4 = L_-$. Showing explicitly the Feynman $i\epsilon$ prescription one has that (\ref{IRmass}) becomes
\begin{equation}
\frac{1}{s}\int d^2L_t\int_{-\infty}^{\infty} dL_+\int_{-\infty}^{\infty}dL_-\frac{1}{(L_t^2+L_+L_-+i\epsilon)(L_t^2+L_+L_-+L_+ +i\epsilon)(L_t^2+L_+L_-+L_-+i\epsilon)}.
\end{equation}

Consider first the $L_-$ integral as an integral along the real
axis in the complex $L_-$ space. There are three simple poles
corresponding to the three propagators. Note that this integral
can be made a contour integral by closing the contour with a big
semi-circle at infinity. This does not affect the value of the
integral as the integrand is cubic in $L^{-1}_-$. Note that the
first two poles are located below (above) the real axis for
$L_+>0$ $(L_+<0)$ while the third pole is below (above) the real
axis for $1+L_+>0$ $(1+L_+<0)$. If all poles are on the same half
plane the integral vanishes. Therefore the region of integration
in $L_+$ is restricted to that where $L_+(1+L_+)<0$. In other
words, $L_+\in (-1,0)$.

Carrying out the integral over $L_-$ by closing the contour in the direction where the third pole is located, the integrals left become
\begin{equation}
\label{collinear}
I_4(s,t)|_{\rm IR} \sim \frac{1}{s}\int d^2L_t \int_{-1}^{0} dL_+ \frac{1}{(L_t^2+(1+L_+)m^2+i\epsilon)(L_t^2+L_+(1+L_+)+i\epsilon)}.
\end{equation}
The integral over $L_t$ is a regular integral over $\mathbb{R}^2$. Using polar coordinates $L_t = (r,\theta)$ and integrating over the angular variable one finds the first IR divergence (regulated by $m^2$). This comes from the $r\sim 0$ region of the integral
\begin{equation}
\int_0 \frac{r dr}{(r^2+(1+L_+)m^2+i\epsilon)} \sim {\rm log}(m^2).
\end{equation}
This is a collinear singularity as $L_t^2\sim 0$ means that $L_-\sim 0$ from the location of the pole and hence $L$ becomes collinear with $k_4$.

Evaluating (\ref{collinear}) explicitly is very easy and gives
\begin{equation}
\int_{-1}^0 \frac{dL_+}{m^2-L_+}{\rm log}\left( \frac{L_+}{m^2} \right).
\end{equation}
Here we see the second source of divergence; the soft singularity around $L_+\sim 0$. Note that this is also regulated by $m^2$ and gives the ${\rm log}^2(m^2)$ behavior as advertised in (\ref{IRmass}).

Now we are ready to see what happens in split signature. Let us use
the same $i\epsilon$ prescription and postpone momentarily the
introduction of the principal value prescription. We have to start
from (\ref{collinear}). The integral over $L_t$ is now over
$\mathbb{R}^{1,1}$. It is natural once again to choose light cone
coordinates $L_t=(\ell_+,\ell_-)$ and write
\begin{equation}
\label{IRone}
I_4(s,t)|_{\rm IR} \sim \frac{1}{s} \int_{-1}^{0}\!\!d\ell_+\int_{-\infty}^{\infty} \!\! d\ell_- \int_{-\infty}^{\infty} \!\! dL_+ \frac{1}{(\ell_+\ell_-+(1+L_+)m^2+i\epsilon)(\ell_+\ell_-+L_+(1+L_+)+i\epsilon)}.
\end{equation}
Considering the $\ell_-$ integral as an integral over the real
axis of the complex $\ell_-$ plane we can repeat the same analysis
as above. There are only two poles and both are located on the
same half plane. Since the integral converges one finds zero as
the answer. Does this mean that loop integrals in split signature
vanish? The answer is yes for generic momenta. As we will see,
loop integrals can have singular support. Here we do not see any
clue of such singular support because we implicitly assumed a
generic point where $k_1\cdot k_4\neq 0$ in order to define the
light cone coordinates. Below we will consider the same integral
under the principal value prescription. We will find that the
integral over $\ell_-$ is non-zero and gives rise to an integral
over $\ell_+$ which is divergent and needs a regulator. This
divergence is the split signature analog of the soft and collinear
singularities found in the lorentzian case.

Up to this point we have considered only the behavior of the box
integral near singularities where three inverse propagators vanish
and therefore it has been enough to study the one-mass triangle
integral. In split signature, there is a new singularity not present
in the lorentzian case. These are points in $L$ where all four
inverse propagators vanish! In order to expose the new isolated
singularities let $L^*$ be one of the two points where all four
inverse propagators vanish. Changing variables to bring $L^*$ to the
origin, i.e., $L \to L^* + L$, one finds,
\begin{eqnarray}
\label{IRsplit}
&& I_4(s,t)|_{\rm IR-split} \sim \delta^4(k_1+k_2+k_3+k_4)\times \\ && \int d^4L \frac{1}{(L^2+2 L^*\cdot L)(L^2+2L\cdot (L^*+k_1))(L^2+2L\cdot (L^*+k_1+k_2))(L^2+2L \cdot (L^*-k_4))}.\nonumber
\end{eqnarray}
which is clearly divergent near $L\sim 0$. A formal theory of the regularization of these IR singularities is out of the scope of this paper. We will instead move on very naively and attempt to use the principal value prescription for the propagators and then transform to twistor space to learn how to tame these singularities.

Also worth mentioning in passing is the fact that in lorentzian signature the IR behavior relates one-loop amplitudes to tree amplitudes~\cite{IRloop}. More precisely, the coefficient of the most singular terms is universal and governed by the tree amplitude, i.e,
\begin{equation}
M^{1\hbox{-}{\rm loop}}_n|_{\rm IR} = M^{\rm tree}_n \times \frac{1}{\epsilon^2}\sum_{i=1}^n (s_{i,i+1})^{-\epsilon}
\end{equation}
In split signature, the behavior of the most singular IR singularities is then controlled by the quadruple cut introduced in \cite{quadruple}!

\subsection{Feynman $i\epsilon$ versus Principal Value}

As mentioned before, the reason for using the Feynman $i\epsilon$ prescription is to ensure physical properties like unitarity. In split signature we do not have such a notion and therefore other prescriptions become available. Let us start once again with the four-particle scalar integral (\ref{four}) rewritten as follows
\begin{equation}
\label{omi}
I_4(s,t) = \int \frac{d^4L_i}{L_i^2}\delta^4(L_{i-1}-L_i+k_i).
\end{equation}
Now we would like to take each propagator and replace Feynman's prescription, $1/(L^2_i + i\epsilon)$, by the principal value, ${\tt p.v.}(1/L_i^2) = 1/2(1/(L^2+i\epsilon) + 1/(L^2-i\epsilon))$.

Using this prescription we find that even after introducing the mass regulator as in (\ref{IRmass}) the loop integral is ill-defined due to the $1/\ell_+$ integral left after the $\ell_-$ integral is performed in (\ref{IRone}).

Here is where we propose to generalize the principal value prescription to regulate these divergences. Note that in lorentzian signature, no $i\epsilon$ prescription can possibly remove the collinear divergence since the integral over the radial part of $L_t$ starts at zero! In this sense, split signature is better behaved than any other signature.

In preparation for the transformation into twistor space, let us
study the scalar integral (\ref{omi}) in yet another
parametrization. Let each loop variable be $L_i = \ell_i+\tau_i
q_i$ with $q_i^2=0$. Let us determine what the principal value
prescription does in this parametrization by first writing the
familiar Feynman prescription\footnote{This lightcone version was
used in \cite{Vaman:2005dt} to relate the BCFW recursion relations
to the largest time equation.}
\begin{equation}
\label{feyn}
\int \frac{d^4L_i}{L_i^2+i\epsilon} = \int d^4\ell_i \delta(\ell_i^2)\int \frac{d\tau_i}{(\tau_i+i\epsilon)}\theta(q_i\cdot \ell_i)
-\int d^4\ell_i \delta(\ell_i^2)\int \frac{d\tau_i}{(\tau_i-i\epsilon)}\theta(-q_i\cdot \ell_i)
\end{equation}
Now it is easy to see what the PV gives by combining (\ref{feyn}) and its complex conjugate
\begin{equation}
\int d^4L_i \;{\tt p.v.}\left( \frac{1}{L_i^2}\right) =\int d^4\ell_i \delta(\ell_i^2)\int d\tau_i\frac{1}{2}\left(\frac{1}{\tau_i+i\epsilon}+\frac{1}{\tau_i-i\epsilon}\right){\rm sgn}(q_i\cdot \ell_i)
\end{equation}
which means that
\begin{equation}
\int d^4L_i \;{\tt p.v.}\left( \frac{1}{L_i^2}\right) =\int d^4\ell_i \delta(\ell_i^2)\int d\tau_i\;{\tt p.v.}\left(\frac{1}{\tau_i}\right){\rm sgn}(q_i\cdot \ell_i)
\end{equation}
Therefore the one-loop integral (\ref{four}) becomes
\begin{equation}
\label{splitfour}
I_4(s,t) = \int \prod_{i=1}^4 d^4\ell_i \delta(\ell_i^2)\int_{-\infty}^{\infty} d\tau_i\;{\tt p.v.}\left(\frac{1}{\tau_i}\right){\rm sgn}(q_i\cdot \ell_i)\delta^4(\ell_{i-1}-\ell_i+k_i(\tau))
\end{equation}
with
\begin{equation}
\label{veci}
k_i(\tau)  =k_i + \tau_{i-1}q_{i-1} -\tau_i q_i.
\end{equation}
Note that the $\tau$ integrations are defined over all the real axis. This is a consequence of working in split signature.

Let us now match the pole we find in $\tau$ with the singularities
found in the previous section. Consider first the singularity at
$\tau_1=\tau_2=\tau_3=\tau_4=0$. At this point, the $L_i$'s in the
original integral become localized at $L^*$ which makes all four
inverse propagators vanish. This is therefore the new singularity
not present in lorentzian signature. Note that the principal value
regulates it completely. This means that an analysis of
(\ref{IRsplit}) similar to the that of (\ref{IRmass}) will show
that the divergence is absent.

We seem to be missing the four other singularities which are also present in the lorentzian case. In order to make those manifest we have to complete the evaluation of (\ref{splitfour}). Before doing this it is simpler to compute the fourier transform of the amplitude. This will reveal that the answer is perfectly well defined. Transforming back into momentum space will give us a hint of where the singularities are and how to regulate them.

\subsection{Four-Particle Amplitude in Twistor Space}

Let us compute the twistor transform of the one-loop four-particle amplitude of gluons $M^{1\hbox{-}{\rm loop}}_{(1^-,2^+,3^-,4^+)}$. This is given by (\ref{ampi}) with $I_4(s,t)$ defined in (\ref{splitfour}). We argued that the one-loop scalar integral $I_4(s,t)$ has IR divergencies. Here we will ignore this issue and transform into twistor space following the same prescriptions used in the rest of this paper at tree level. We will find that the twistor space answer is finite and perfectly well defined.

Let us start by computing the prefactor of the scalar integral,
\begin{equation}
\label{pref}
{\cal M}^{\rm tree}_{(1^-,2^+,3^-,4^+)}st = \frac{\langle 1~3\rangle^4}{\langle 1~2\rangle\langle 2~3\rangle\langle 3~4\rangle\langle 4~1\rangle}st = \langle 1~3\rangle^2[2~4]^2.
\end{equation}

The transformation into twistor space is defined as follows
\begin{equation}
\label{timo}
M^{1\hbox{-}{\rm loop}}(W,Z) = \int d^2\tilde\lambda_1 e^{i\mu_1^{\dot a}\tilde\lambda_{1,\dot a}}\int d^2\lambda_2 e^{i \tilde\mu_2^a\lambda_{2,a}}\int d^2\tilde\lambda_3 e^{i\mu_3^{\dot a}\tilde\lambda_{3,\dot a}}\int d^2\lambda_4 e^{i\tilde\mu_4^a \lambda_{4,a}}M^{1\hbox{-}{\rm loop}}.
\end{equation}
here by $(W,Z)$ we mean the specific combination $(Z_1,W_2,Z_3,W_4)$.
Note that the prefactor (\ref{pref}) comes out of the integral and we are left with the computation of the twistor form of the one-loop scalar integral.

Motivated by the way the BCFW deformation parameter $\tau$ was
pulled out of the product of the amplitudes we choose reference
vectors $q_i$ so that all $\tau_i$'s can be pulled out in a
similar way. Let
\begin{equation}
\label{choice}
q_1 =\lambda_1\tilde\lambda_2, \; q_2 = \lambda_3\tilde\lambda_2, \; q_3 = \lambda_3\tilde\lambda_4, \; q_4 = \lambda_1\tilde\lambda_4.
\end{equation}
with this choice the scalar integral becomes a function of $k_i(\tau)$ defined in (\ref{veci}),
\begin{eqnarray}
k_1(\tau) &=& \lambda_1 (\tilde\lambda_1+\tau_4\tilde\lambda_4-\tau_1\tilde\lambda_2), \nonumber \\ k_2(\tau) & = & (\lambda_2+\tau_1\lambda_1-\tau_2\lambda_3)\tilde\lambda_2, \nonumber \\ k_3(\tau) & = & \lambda_3 (\tilde\lambda_3 + \tau_2\tilde\lambda_2 - \tau_3\tilde\lambda_4 ), \nonumber \\ k_4(\tau) &=& (\lambda_4+\tau_3\lambda_3-\tau_4\lambda_1)\tilde\lambda_4.
\end{eqnarray}
It is important to mention that this choice of reference spinor can only be made in split signature as in lorentzian signature the vectors $q_i$ as defined in (\ref{choice}) are complex and take the vectors $L_i$ out of the contour of integration.

The next step is easy to guess. Let us exchange the order of integrations and pull the $\tau$ integrals in (\ref{splitfour}) out of the fourier transforms in (\ref{timo}). Once this is done, a simple change of integration variables of the form $\tilde\lambda_1\to \tilde\lambda_1-\tau_4\tilde\lambda_4+\tau_1\tilde\lambda_2$, etc., removes all $\tau$ dependence from the function to be transformed into twistor space. Just as in the BCFW computation in section 4.2 we find that the $\tau$ integrals are of the form
\begin{equation}
\int_{-\infty}^{\infty}\frac{d\tau}{\tau}e^{i\tau Z\cdot W} = {\rm sgn}(Z\cdot W)
\end{equation}
for some $Z$ and $W$ of the external particles. Here we have used that the distribution $1/\tau$ is defined using the principal value prescription in (\ref{splitfour}).

Combining all these steps we find
\begin{equation}
M^{1\hbox{-}{\rm loop}}(W,Z) = M^{\rm tree}(W,Z) \langle 1~3\rangle^2[2~4]^2 K
\end{equation}
with $M^{\rm tree}(W,Z)$ the tree level amplitude in twistor space ${\rm sgn}(Z_1\cdot W_4){\rm sgn}(Z_1\cdot W_2){\rm sgn}(Z_3\cdot W_2){\rm sgn}(Z_3\cdot W_4)$ and
\begin{eqnarray}
& K = \int d^2\tilde\lambda_1 e^{i\mu_1^{\dot a}\tilde\lambda_{1,\dot a}}\int d^2\lambda_2 e^{i \tilde\mu_2^a\lambda_{2,a}}\int d^2\tilde\lambda_3 e^{i\mu_3^{\dot a}\tilde\lambda_{3,\dot a}}\int d^2\lambda_4 e^{i\tilde\mu_4^a \lambda_{4,a}}\times \nonumber\\ & \prod_{i=1}^4\int d^4\ell_i\delta(\ell_i^2)\delta^4(\ell_{i-1}-\ell_{i}-\lambda_i\tilde\lambda_i){\rm sgn}(q_i\cdot \ell_i).
\end{eqnarray}

The integrals in the second line of the definition of $K$ are nothing but the quadruple cut of a one-loop scalar integral with external momenta given by $k_1$, $k_2$, $k_3$ and $k_4$! In other words, the delta functions are enough to localize all $\ell_i$'s and the answer is simply a jacobian. There are two solutions to the equations imposed by the delta functions. The jacobian in both cases is given by $1/|st|$. The absolute value is due to the fact that we are working with real variables. Finally one has to evaluate the product of ${\rm sgn}(q_i\cdot \ell_i)$ in the two solutions $\ell_i^*$. It turns out that on one solution each factor vanishes while on the second one the product of all four factors gives ${\rm sgn}(st)$. Combining these results one finds
\begin{eqnarray}
&& M^{1\hbox{-}{\rm loop}}(W,Z) = M^{\rm tree}(W,Z) \times \\ && \int d^2\tilde\lambda_1 e^{i\mu_1^{\dot a}\tilde\lambda_{1,\dot a}}\int d^2\lambda_2 e^{i \tilde\mu_2^a\lambda_{2,a}}\int d^2\tilde\lambda_3 e^{i\mu_3^{\dot a}\tilde\lambda_{3,\dot a}}\int d^2\lambda_4 e^{i\tilde\mu_4^a \lambda_{4,a}}\frac{\langle 1~3\rangle^2[2~4]^2}{st}\delta^4(k_1+k_2+k_3+k_4).\nonumber
\end{eqnarray}
Note the amusing fact that the integrand, $\langle 1~3\rangle^2[2~4]^2/st \delta^4(k_1+k_2+k_3+k_4)$, is nothing but ${\cal M}^{\rm tree}\delta^4(k_1+k_2+k_3+k_4) = M^{\rm tree}$ which means that
\begin{equation}
\label{fino}
M^{1\hbox{-}{\rm loop}}(W,Z) = (M^{\rm tree}(W,Z))^2 = 1.
\end{equation}
This is our final result.

This formula can be drawn as a Hodges diagram:
\be
\includegraphics[scale=0.8]{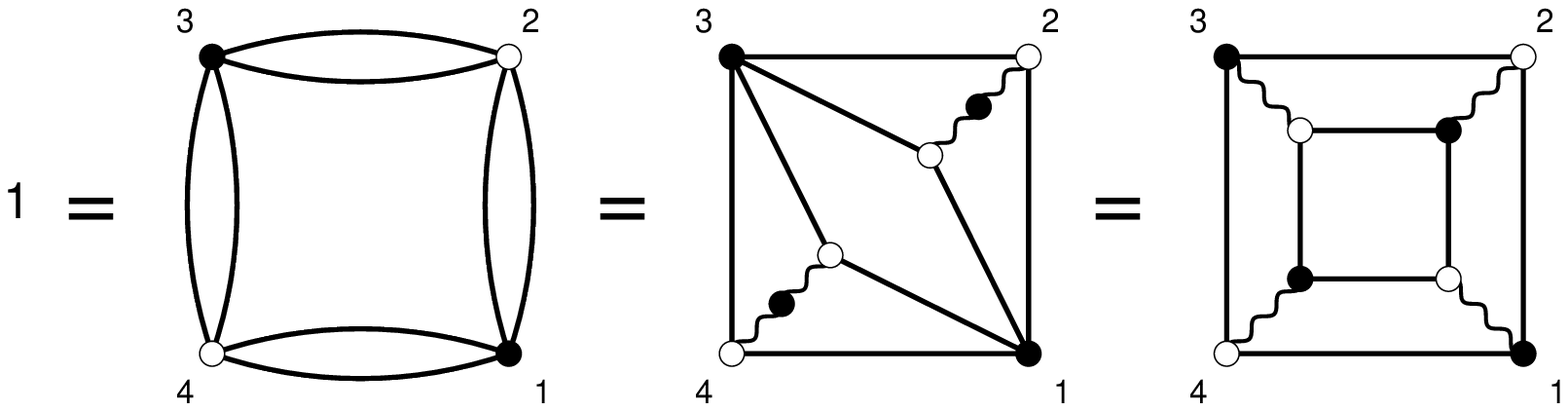} \nonumber
\ee
The squashed figure can be blown up by twistor transforming twice two diagonally opposed vertices. Finally, using the square identity this can be brought to the form of a Hodges diagram with the topology of an annulus. This is very suggestive, especially given that tree amplitudes have the topology of a disk! We postpone the exploration of loop level Hodges diagrams for future work.

\subsection{Four-Particle Amplitude Back Into Momentum Space}

One might wonder how is it possible that starting with a divergent integral we found such a simple and well defined answer as $M^{1\hbox{-}{\rm loop}}(Z_1,W_2,Z_3,W_4)=1$ in twistor space. One might say that the step of exchanging the integrations by pulling the $\tau$ integrals out of the fourier transforms is not valid. As we will see below this is not the case as the kind of singularities left are actually equivalent to the singularities one has to regulate in the final step that led to equation (\ref{fino}), i.e., in defining the fourier transform of $M^{\rm tree}$. Note that from the one-loop point of view there is no reason to use the principal value prescription except in the $\tau$ integrals. However, the fact that the computation involves precisely the fourier transform of the tree level amplitude led us to use the principal value once again to render the transform well defined.

The natural question is what this prescription corresponds to in a direct evaluation of the integrals in momentum space. In order to answer this question let us complete the evaluation of $I_4(s,t)$ given in (\ref{splitfour}). Choosing the same reference spinors as in (\ref{choice}) one finds
\begin{equation}
I_4(s,t) = \int_{-\infty}^{\infty} \frac{d\tau_i}{\tau_i} \frac{1}{(k_1(\tau)+k_2(\tau))^2(k_2(\tau)+k_3(\tau))^2}\delta^4(k_1+k_2+k_3+k_4).
\end{equation}
Here we used again that the product of ${\rm sgn}(q_i\cdot\ell_i^*)$ vanishes in one solution of the quadruple cut equations while it gives ${\rm sgn}((k_1(\tau)+k_2(\tau))^2(k_2(\tau)+k_3(\tau))^2)$ in the other. It is also easy to check that $k_1+k_2+k_3+k_4 = k_1(\tau) +k_2(\tau)+k_3(\tau)+k_4(\tau)$.

Simplifying further one arrives at a remarkably simple result
\begin{equation}
I_4 = \frac{1}{\langle 1~3\rangle^2[2~4]^2}\int_{-\infty}^{\infty} \frac{d\tau_i}{\tau_i(\tau_i-A_i)}\delta^4(k_1+k_2+k_3+k_4)
\end{equation}
with
\begin{equation}
A_1 = \frac{[1~4]}{[2~4]} = -\frac{\langle 2~3\rangle}{\langle 1~3\rangle}, \; A_2 = -\frac{\langle 1~2\rangle}{\langle 1~3\rangle} = \frac{[4~3]}{[4~2]}, \; A_3 = -\frac{\langle 1~4\rangle}{\langle 1~3\rangle} = \frac{[2~3]}{[2~4]}, \; A_4 =\frac{[1~2]}{[4~2]} = -\frac{\langle 4~3\rangle}{\langle 1~3\rangle}.
\end{equation}
The second equalities are a consequence of momentum conservation.

Now we can complete the identification of singularities we started
after the definition of (\ref{splitfour}). Recall that we had
identified the pole at $\tau_1=\tau_2=\tau_3=\tau_4=0$ with the
new singularity in split signature where all four inverse
propagators vanish. This is completely regulated using principal
value.  Consider singularities where three $\tau_i$ vanish and one
is equal to $A_i$. There are four such singularities. Note that
the only piece left ill-defined is the region near $\tau_i = A_i$.
This matches the $\ell_+$ integral left in (\ref{splitfour}). This
means that these four singularities correspond to the four
singularities in lorentzian signature. Other combinations of
$\tau_i=0$ and $\tau_i=A_i$ vanish on the support of the delta
functions.

Finally, let us show that by also using principal value to define $1/(\tau_i-A_i)$ one reproduces the answer for the amplitude obtained from twistor space. Using this prescription one might think that each $\tau$ integral is zero. This is indeed the case for any $A_i\neq 0$. Therefore one is left with computing the integral at $A_i=0$. Treating the integral as a distribution, multiplying by a test function $\phi(A_i)$ and integrating over $A_i$ one finds that the answer is $\phi(0)$. This can be proven by writing $\phi(A_i)$ in its Fourier representation and then carrying out the $A_i$ and $\tau_i$ integrals. This shows that each $\tau_i$ integral gives a delta function $\delta(A_i)$.

Using this the scalar one-loop integral becomes
\begin{equation}
I_4 = \frac{1}{\langle 1~3\rangle^2[2~4]^2}\delta(A_1)\delta(A_2)\delta(A_3)\delta(A_4)\delta^4(k_1+k_2+k_3+k_4)
\end{equation}

Finally, we can multiply by the prefactor (\ref{pref}) to obtain the amplitude. It is very useful to choose the new representation for the delta function found in section 3.2,
\begin{eqnarray}
&& M^{1\hbox{-}{\rm loop}} = \delta(A_1)\delta(A_2)\delta(A_3)\delta(A_4) \int dc_{21}dc_{41}dc_{23}dc_{43} \delta^2(\tilde\lambda_1-c_{21}\tilde\lambda_2-c_{41}\tilde\lambda_4)\times \\ &&
\delta^2(\lambda_2 -c_{21}\lambda_1-c_{23}\lambda_3) \delta^2(\tilde\lambda_3-c_{23}\tilde\lambda_2-c_{43}\tilde\lambda_4)
\delta^2(\lambda_4-c_{41}\lambda_1-c_{43}\lambda_3).\nonumber
\end{eqnarray}
Note that $A_i$'s are precisely the values of the $c$'s on the support of the delta functions, therefore we can introduce the delta functions into the integrals to get
\begin{eqnarray}
 &&M^{1\hbox{-}{\rm loop}} = \int dc_{21}dc_{41}dc_{23}dc_{43}\delta(c_{21})\delta(c_{41})\delta(c_{43})\delta(c_{21}) \delta^2(\tilde\lambda_1-c_{21}\tilde\lambda_2-c_{41}\tilde\lambda_4)\times  \\ &&
\delta^2(\lambda_2 -c_{21}\lambda_1-c_{23}\lambda_3) \delta^2(\tilde\lambda_3-c_{23}\tilde\lambda_2-c_{43}\tilde\lambda_4)
\delta^2(\lambda_4-c_{41}\lambda_1-c_{43}\lambda_3).\nonumber
\end{eqnarray}
This is nothing but the link representation of the tree-level amplitude where in the integrand we have replaced each factor of $1/c_{ij}$ by $\delta(c_{ij})$.

In this form, finding the twistor transform of the loop amplitude is straightforward and gives
\begin{equation}
M^{1\hbox{-}{\rm loop}}(Z_1,W_2,Z_3,W_4) = 1.
\end{equation}
as expected.

\subsubsection{A Subtlety}

Let us go back to a subtlety in our computation. The question is the validity of the choice of reference vectors made in (\ref{choice}). On the support of the delta functions, we find that $s=t=0$ and therefore $u=0$. The latter equation gives $\langle 1~3\rangle [1~3] = \langle 2~4\rangle [2~4] = 0$. Our choice of reference vectors is valid where $[1~3] = \langle 2~4\rangle = 0$ while $\langle 1~3\rangle\neq 0 \neq [2~4]$. There is a second choice of reference vectors which is natural and it is obtained from the first by exchanging $\lambda$'s with $\tilde\lambda$'s in (\ref{choice}). This choice is valid when  $[2~4] = \langle 1~3\rangle = 0$ while $\langle 2~4\rangle\neq 0 \neq [1~3]$. Using this choice the first solution to the quadruple cut equations gives zero and the second one contributes.

Summarizing, a form of the scalar integral valid for any momenta is given by simply adding the two choices as they do not share the same support. Let us write the final form as
\begin{equation}
I_4 = \left( \frac{1}{\langle 1~3\rangle^2[2~4]^2}\delta(A_1)\delta(A_2)\delta(A_3)\delta(A_4) + \{\lambda\leftrightarrow \tilde\lambda\} \right)\delta^4(k_1+k_2+k_3+k_4).
\end{equation}

Using this to compute the full one-loop amplitude with helicities $M(-+-+)$ the prefactor $M^{\rm tree}st = \langle 1~3\rangle^2[2~4]^2$ vanishes on the support of the second term. If we consider instead $M(+-+-)$ then the prefactor vanishes in the first term and contributes in the second. It is interesting to note that resemblance of this form to that of the three-particle amplitude in the full ${\cal N}=4$ SYM which also possesses two terms. In fact, we can very well use the full supersymmetric four-particle amplitude and note that the only components that contribute are the ones with alternating helicity. This is due to the fact that the zeroes coming from the $st$ factor are not canceled by poles in $M^{\rm tree}$.

The ${\cal N}=4$ supersymmetric formula is given by
\begin{equation}
M^{1\hbox{-}{\rm loop}} = \frac{\delta^8\left(\sum_{i=1}^4\lambda_i\tilde\eta_i^I\right)}{\langle 1~2\rangle\langle 2~3\rangle\langle 3~4\rangle\langle 4~1\rangle} \times st \times I_4(s,t)
\end{equation}
Expanding this in powers of $\eta$, the individual helicity amplitudes are either ``1" or ``0".

\section{Outlook}

In this note we have clearly only scratched the surface of what appears to be a marvelous structure underlying scattering amplitudes in twistor space. Our ``ambidextrous" transformation to twistor space, together with its natural marriage with the BCFW formalism, has allowed us to use quantum field theory itself as our guide to discovering the nature of its structure in twistor space. We will develop many of these themes further in \cite{HolS}, which will lead us to a completely different picture for computing scattering amplitudes at tree level than given by the BCFW formalism, that we strongly suspect is connected with a maximally holographic description of tree amplitudes that makes all the symmetries of the theory manifest but completely obscures space-time locality.

We have already given one holographic definition of ${\cal N} = 4$ SYM and ${\cal N} = 8$ SUGRA at tree level, by the quadratic equations (\ref{quadYM}) and (\ref{quadGRA}). The S-Matrix generating functionals are the analogs of the very familiar effective action $\Gamma(\phi)$ in quantum field theory. It is therefore tempting to find the analog of the effective potential $V(\phi)$, and more generally, to find solutions to these equations directly without doing a perturbative expansion. It is also tempting to ask whether there is a natural deformation of these equations, with the parameter $\hbar$, that can automatically encode loop corrections in a simple way.

There are also a large number of open avenues of exploration in further developing the twistorial formalism. Most pressingly, while we have benefited greatly from being able to do concrete computations in $(2,2)$ signature, we would really like to be able to recast these as contour integrals in complexified twistor space; our $(2,2)$ formalism should be thought of as defining a ``good" contour of integration, but it should be possible to deform this contour to define the theory in $(3,1)$ signature. This should be especially helpful at loop level, where the IR divergences with their important physical interpretation must emerge.

Finally, the over-arching question raised by our work is an obvious one: we have argued that there may be a theory  naturally living in (2,2) signature, that plays the same role for defining (3,1) scattering amplitudes as Euclidean quantum field theory plays for defining (3,1) correlation functions. We have a nice interpretation for the correlation functions in Euclidean space. But in tandem with trying to characterize the putative (2,2) dual theory, we should also ask: what is the ``observable" in (2,2) signature that corresponds to the scattering amplitudes?

\section*{Acknowledgments}

We thank Fernando Alday, Ruth Britto, Henriette Elvang, and
especially Juan Maldacena and Edward Witten for many stimulating
comments. We also thank Andrew Hodges, Lionel Mason and David
Skinner for several days of extremely enjoyable and productive
discussions of our related works in progress. F.C. is also
grateful to the Institute for Advanced Study for hospitality
during the origination of this work. N-A.H., C.-C. and J.K.
similarly thank the Perimeter Institute for its hospitality.
N.A.-H. is supported by the DOE under grant DE-FG02-91ER40654,
F.C. was supported in part by the NSERC of Canada and MEDT of
Ontario, and J.K. is supported by a Hertz foundation fellowship
and an NSF fellowship.

\end{document}